\documentclass[letterpaper]{article}

\usepackage{aaai19}  
\usepackage{times}  
\usepackage{helvet}  
\usepackage{courier} 
\usepackage[hyphens]{url}
\usepackage{graphicx}
\usepackage{graphicx}  
\usepackage{booktabs}
\usepackage{epsfig}
\usepackage{amsmath}
\usepackage{amssymb}
\usepackage{bm}
\usepackage{comment}
\usepackage{epstopdf}  
\usepackage{color}
\usepackage{algorithmic}
\usepackage{layout}
\usepackage{subcaption}
\usepackage{multirow}
\usepackage{xcolor}
\usepackage{booktabs}
\usepackage{array}
\usepackage[skip=10pt]{caption}  
\usepackage{soul}

\newcolumntype{L}[1]{>{\raggedright\let\newline\\\arraybackslash\hspace{0pt}}m{#1}}
\newcolumntype{C}[1]{>{\centering\let\newline\\\arraybackslash\hspace{0pt}}m{#1}}
\newcolumntype{R}[1]{>{\raggedleft\let\newline\\\arraybackslash\hspace{0pt}}m{#1}}

\newcommand{\todo}[1]{\textcolor{red}{#1}}
\DeclareMathOperator*{\argmax}{arg\,max}
\DeclareMathOperator*{\argmin}{arg\,min}

\title{Scenario Generalization of Data-driven Imitation Models in Crowd Simulation} 

\author{
Gang Qiao$^{1}$, Honglu Zhou$^{1}$, Mubbasir Kapadia$^{1}$, Sejong Yoon$^{2}$, Vladimir Pavlovic$^{1}$ \\
	$^{1}$Rutgers University, $^{2}$The College of New Jersey \\
    \{gq19, mk1353, vladimir\}@cs.rutgers.edu, hz289@scarletmail.rutgers.edu, yoons@tcnj.edu
}

\begin{document}

\maketitle

\begin{abstract}  
Crowd simulation, the study of the movement of multiple agents in complex environments, presents a unique application domain for machine learning. One challenge in crowd simulation is to imitate the movement of expert agents in highly dense crowds. An imitation model could substitute an expert agent if the model behaves as good as the expert. This will bring many exciting applications. However, we believe no prior studies have considered the critical question of how training data and training methods affect imitators when these models are applied to novel scenarios. In this work, a general imitation model is represented by applying either the Behavior Cloning (BC) training method or a more sophisticated Generative Adversarial Imitation Learning (GAIL) method, on three typical types of data domains: standard benchmarks for evaluating crowd models, random sampling of state-action pairs, and egocentric scenarios that capture local interactions. Simulated results suggest that (i) simpler training methods are overall better than more complex training methods, (ii) training samples with diverse agent-agent and agent-obstacle interactions are beneficial for reducing collisions when the trained models are applied to new scenarios. We additionally evaluated our models in their ability to imitate real world crowd trajectories observed from surveillance videos. Our findings indicate that models trained on representative scenarios generalize to new, unseen situations observed in real human crowds. 
\end{abstract}


\section{Introduction}

Imitating the movement of a goal-directed expert agent in a complex scenario, involving obstacles and other agents, has recently received attention from machine learning community. Researchers aim to create data-driven models to predict the next movement decision (velocity) of an agent given current state (local observation on environment and neighboring agents), by imitating the demonstrated crowd movement of an expert. A good imitator could substitute the expert, with potentialities in some applications. For instance, we may want to imitate the controlling signals (steering angle, acceleration, etc.) demonstrated by a real person steering a vehicle in parallel parking scenarios, whose decisions are based on the person's successive local observations, and then replace the human efforts with the imitator to provide controlling signals given the observations of a camera equipped on the vehicle in new parallel parking scenarios.

In this paper, the term ``scenario'' refers to the configuration of environment obstacles as well as the tasks (initial positions and destination positions) for all involved agents. Agents may have different destination positions. Existing works, e.g.,~\cite{qiao2018role}, train and test models over the same environment with only initial and goal positions and the number of agents varied, or over environments with small obstacle adjustment ~\cite{long2017deep}. To the best of our knowledge, no prior studies have considered the critical question of how training data and training paradigms affect imitation models when these models are generalized to substantially different scenarios.

The generalization ability of an imitator to new scenarios is subtly but essentially different from the regular generalization ability of a model, in three aspects. (1) For regular generalization, the model has full knowledge about a scenario such as the initial/destination positions of all agents and the positions of all obstacles, while in scenario generalization, each agent assigned with an imitator may only know its own destination and make a decision based on its own partial observation, without knowing the destinations and observations of other agents. (2) For regular generalization, test samples are usually isolated: a previous test sample does not influence the next test sample. In contrast, in scenario generalization all agents (each agent is equipped with an imitator) move step by step and synchronously, during which the previous observation and decision of an agent influence its next observation and decision successively. (3) Instead of measuring on isolated state-action pairs in regular generalization, measurement in scenario generalization is on the overall generated trajectories with multiple metrics, and some of them might be mutually balanced.

Unlike most previous works that focus on improving a specific expert model, or imitating an expert model for a specific behavior/in a specific scenario, our main goal is to investigate the effect of training paradigm and training data on the scenario generalization ability of an imitator, by comparing the combinations of representative training paradigms and representative data domains. Specifically, two training paradigms are studied: (1) \textbf{Behavior Cloning} (BC): an approximation of maximum likelihood estimation by fitting a neural network regressor, capable of representing many classic regressors (support vector regressor, random forest, etc.) and (2) \textbf{Reinforcement Learning} (RL): a Markov decision process solved by Generative Adversarial Imitation Learning (GAIL)~\cite{ho2016generative}, leading to a solution that is theoretically equivalent to any two-step reward estimation followed by policy search procedures. Although only two paradigms are studied, encompassing two distinct families of training approaches, the former focusing on imitating simple reactive behaviors, while the latter considering the impact of local actions on accumulated outcomes, they are generic modeling approaches and represent most data-driven models in crowd simulation.

In addition to training paradigms, three data domains are developed: (1) a set of six standard scenarios which serve as benchmarks for evaluating crowd simulation,  (2) a random sampling of inter-agent interactions at a single time step, and (3) a set of representative scenarios to capture inter-agent and agent-obstacle interactions during the overall navigation procedure. These data domains span the spectrum of a few but complex and crowded scenarios, to many random discrete snapshots for the immediate response of a model to inter-agent interactions, and a large number of sampling of small-scale, but general interaction situations that individuals encounter.

Combinations of training paradigms and data domains are systematically evaluated in the ability to emulate expert trajectories while avoiding collisions with other agents and environment obstacles in substantially new scenarios. Our empirical results suggest that (i) a simpler training method is better than a more complex training method, (ii) training samples with diverse agent-agent and agent-obstacle interactions are beneficial for reducing collisions when the trained models are applied to new scenarios.

We additionally evaluated all five models in their ability to imitate real world crowd trajectories observed from surveillance videos. Results indicate that models trained on representative scenarios generalize to new, unseen situations observed in real human crowds.


\begin{figure*}[t]
 \centering
  \includegraphics[width=\textwidth,height=5cm]{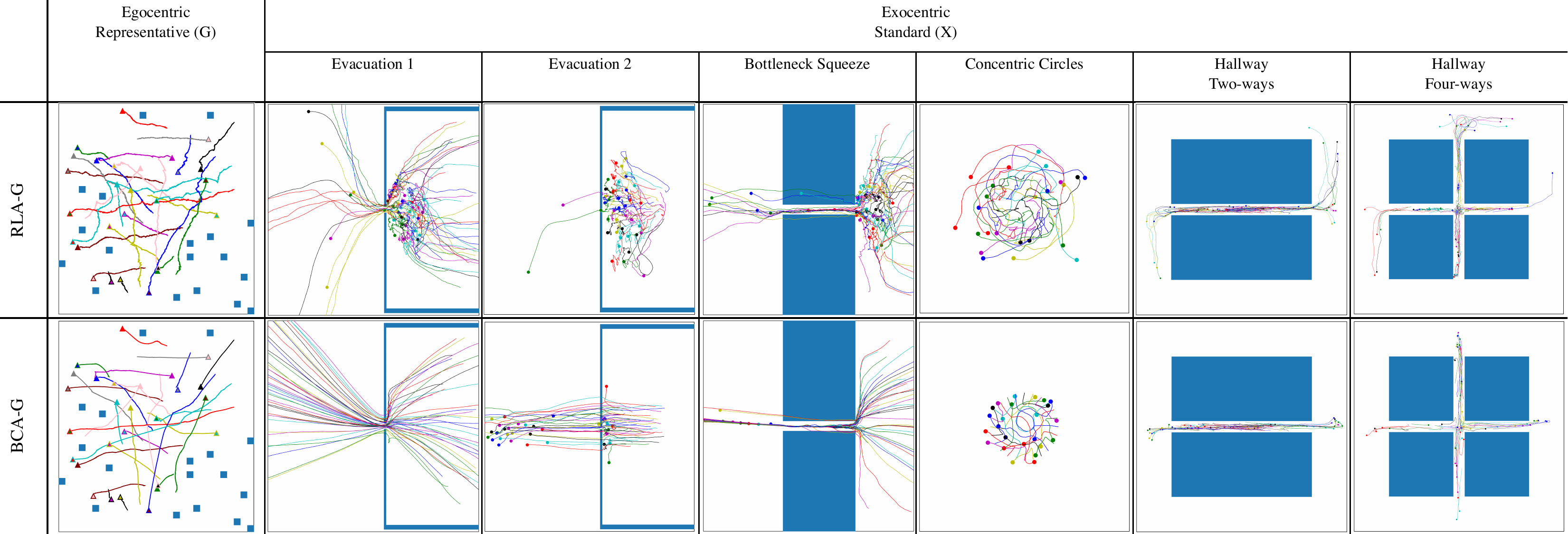}
  \caption{Visualization of trajectories of \texttt{RLA-G} and \texttt{BCA-G} generalized to egocentric representative and exocentric standard scenarios. The lower AO metric of \texttt{RLA-G} than that of \texttt{BCA-G} results from the fact that much fewer \texttt{RLA-G} agents can avoid obstacles and reach destinations.}
  \label{fig:comparative_trajs_on_F} 
\end{figure*}

\section{Prior work}
\label{sec2}

Crowd simulation and analysis are paramount examples of distributed AI modeling, with application across a variety of domains including computer graphics, crowd tracking, crowd trajectory estimation and optimization ~\cite{ali2013modeling,junior2010crowd,kapadia2015virtual,qiao2018role,Liu:2017:CRE:3136457.3136474,Lee:2018:CSD:3274247.3274510,cheng2018}. We provide a brief summary of the most related literature below. 


\subsection{Crowd Simulation Approach}
Methods in this approach rely on pre-determined physical, social or geometric rules or computational procedures to decide a velocity for an agent to execute in the next time duration ~\cite{vicsek1995novel,karamouzas2014universal,knob2019optimal,kim2012interactive,kim2013velocity,ren2017group}, and hence they are not data-driven models. In social force method ~\cite{helbing1995social}, an agent is simultaneously attracted by its goal and repelled by other agents and obstacles. Each force obeys the gravitation-like inverse-square law, and the composition of all forces of an agent determines the acceleration of that agent. Geometric methods such as velocity obstacles ~\cite{fiorini1998motion} define a geometrical cone in the relative velocity space, inside which a collision will occur. Extensions to this work ~\cite{van2011reciprocal} define the set of collision-avoiding velocities and induce Optimal Reciprocal Collision Avoidance (ORCA) that provides a sufficient condition for collision avoidance if agents are not densely packed.

\subsection{Behavior Cloning (BC) Approach}
This approach view state-action $(s,a)$ pairs as independent samples and use these samples to fit a regression model based on maximum likelihood estimation (MLE). Thus, models ~\cite{long2017deep,qiao2018role,torabi2018behavioral} within this approach are data-driven models. If the regression model is represented by a neural network (NN), it stands for a general function and covers many traditional learning models. \cite{long2017deep} randomly places neighboring agents around a reference agent and randomly samples the current velocities for all agents. Given a preferred velocity for the reference agent, they use ORCA~\cite{van2011reciprocal} to produce the corresponding action (velocity) for the reference agent in that state. Such a uniform sampling over state space yields a sufficient amount of state-action pairs to fit an NN model. Similarly, ~\cite{qiao2018role} simulates the social force model to collect expert trajectories, and treat state-action pairs from the expert trajectories as independent samples to fit an NN model. However, the trained model is used to provide a velocity prior used for trajectory interpolation, where the actions of individual agents become seemingly decoupled from each other, leading to a computationally efficient solution.

\subsection{Reinforcement Learning (RL) Approach}
RL methods ~\cite{ziebart2008maximum,finn2016connection,arora2018survey,schulman2015trust,pautrat2018bayesian,ho2016generative,kuefler2017imitating} alternate between sampling trajectories with a policy model in an environment and updating the policy model based on reward signal. The goal is to maximize the expected accumulated reward by balancing environment exploration and reward exploitation. ~\cite{torrey2010crowd} introduces RL to crowd simulation and proposes several new challenges when it scales from  single-agent to multi-agent setting. A recent work presents an agent-based, RL navigation method that learns a single unified policy to be applicable to several scenarios and settings, without considering environmental obstacles~\cite{Lee:2018:CSD:3274247.3274510}. Some other works ~\cite{Casadiego2014} also use RL to approach the problem of data-driven trajectories learning~\cite{cheng2018} in crowd simulation.  

The reward function in RL is either human-defined ~\cite{long2018towards}, or learned with inverse-RL (IRL) methods~\cite{ziebart2008maximum,finn2016connection,arora2018survey}. For fair comparison, we consider only data-driven models and thus the reward function is estimated via IRL from demonstrated expert trajectories.

~\cite{finn2016connection} proposes guided cost learning for IRL, which alternates between (1) estimating the partition function (so as to search for the current optimal parameter point) by sampling the proposal distribution, and (2) optimizing the proposal distribution to reduce the variance of the partition function. Given estimated reward function, ~\cite{schulman2015trust} proposes to optimize the policy by searching at each iteration within a region centered at previously estimated parameter point, which could be considered as KL-constrained natural gradient ascend. Recently, ~\cite{ho2016generative} proposes generative adversarial imitation learning (GAIL), an imitator of demonstration. It is model-free, without the need to estimate the dynamics explicitly. More importantly, they proved that any two-step reward estimation and policy optimization procedures (IRL-RL) are equivalent to the one-step adversarial learning. Thus GAIL covers most traditional data-driven RL methods, avoiding us the need to develop a specific RL model. We will describe this training paradigm in detail in the following section, and apply it within the context of multi-agent goal-directed collision avoidance.

\subsection{Comparison of Three Approaches}
The three categories of approaches have their own characteristics, which make them complementary to others. (1) Some methods describe certain movement knowledge of physical particles, geometric objects, animals or humans, and represent the knowledge explicitly for making velocity decisions in crowd simulation, rather than focusing on imitating/learning implicit knowledge from demonstrated data. (2) Provided with expert trajectories, BC suffers from the well-known compounding error problem ~\cite{ross2010efficient}. That is, when BC's decision deviates a little from the expert's decision, the next state would be less represented in expert trajectories, leading to further deviation from the expert decisions. When such error accumulates, it might end up with invalid situation (e.g., off-road driving). (3) RL methods are much more sophisticated in training compared with BC. (4) One can anticipate that the physics-based approach has the best scenario generalization ability, followed by BC, while RL have the least scenario generalization ability. This might be explained by Occam's razor law: physics-based methods follow a few rules or computational procedures, BC learn independently from $(s,a)$ pairs, while RL explore and learn from the same environment repeatedly.

Despite these insights, it is still not clear to which extent the data-driven models differ from each other, in the sense of generalization capacity to new scenarios. Therefore, we specifically seek to determine what training paradigm / training data is the most suitable for developing generalizable models for multi-agent goal-directed collision avoidance. Considering the above-mentioned characteristics of the three approaches, we use physics-based methods to generate different types of expert trajectories and utilize these trajectories for training with BC/RL approaches, followed by comparing the scenario generalization capacities of those trained models.

\section{Problem Formulation}
\label{sec3}
Let $\mathcal{S}$ and $\mathcal{A}$ be the state and the action space, respectively, of an agent given an environment $\mathcal{E}$. Let $s_{t} \in \mathcal{S}$ denote the state of an agent at time $t$, where $t$ is the discrete step index with $t = 0,1..T$ and $T$ is the maximal number of steps. An agent's state typically includes what the agent locally observes about the world around itself, and may also incorporate some guidance signals received from external sources. Let $a_{t} \in \mathcal{A}$ denote the action of the agent at time $t$, determined by the agent's policy function (decision-making function) based on $s_{t}$. That is, $a_{t} =\pi(s_{t})$, with $\pi(\cdot)$ representing the policy function adopted by the agent. The action could be high-dimensional controlling signal (steering angle, acceleration, etc.), but in its simplest form, it may represent the velocity that will take the agent to a new position, leading to a new local observation of the world. At each step $t$, assume the next state $s_{t+1}$ of an agent depends only on its current state $s_{t}$ and current action $a_{t}$. For comparison, we further assume all agents are homogeneous, i.e., they utilize the same policy $\pi$ for their execution, however no agent knows what policies other agents adopt. Therefore, the dynamics, $s_{t+1} \sim P(\cdot | s_{t}, a_{t})$, is probabilistic due to partial observation of the agent and unknowing about other agents' decisions at step t. Furthermore, a state-action pair $(s_{t}, \pi(s_{t}))$ can be evaluated by a cost function associated with the world system: $c(s_{t}, \pi(s_{t}))=r_{t}$, where $r_{t}\in R$ is a reward value for the action the policy decides based on $s_{t}$. For instance, the cost function may evaluate a lower reward if executing $a_{t}$ incurs agent-agent/agent-obstacle collisions and a higher reward otherwise.

Given the above definitions, the problem can be formulated as a Markov Decision Process (MDP). For a given cost function $c(\cdot,\cdot)$, the goal is to find $\pi^*$ that maximizes the accumulated rewards along the expected trajectory:
\begin{align}
\label{eq:expected_cost}
\pi^* = \argmax_{\pi} \mathbb{E}_{\pi}\left[ \sum_{t=1}^{T} \gamma_{t} c(s_{t}, a_{t}) \right],
\end{align}
where $\gamma_{t} \in [0, 1)$ denotes the discount factor.

One issue is that the cost function is usually unknown or implicit, and the demonstrated expert trajectories also conceal the reward signals. In other words, the demonstrated expert trajectories are $\{(s_{0}, a_{0}, s_{1}, a_{1},..., s_{T}, a_{T})\}$, not $\{(s_{0}, a_{0}, r_{0}, s_{1}, a_{1}, r_{1},...,s_{T}, a_{T}, r_{T})\}$. Another important issue is that neither the stochastic dynamics nor the expert policy $\pi_{E}$ is known, stemming from the complex nature of the crowd simulation task. Typically, there are four ways to handle these challenges: (1) use IRL to estimate a cost function that favors the expert trajectories with high accumulated rewards (in the following, we denote the cost function estimated from expert trajectories as c*), (2) estimate the dynamics from data, (3) use RL to estimate $\pi^{*}$ to mimic the expert policy $\pi_{E}$ using the IRL-found cost function $c^{*}$, and (4) use BC to directly estimate $\pi^{*}$ from the expert trajectories. We focus on (3) and (4) in this work.

\section{Behavior Cloning Agents}
\label{sec4}
Behavior cloning methods could be viewed as a special case of the formulation in Eq.~\ref{eq:expected_cost}: a reduction when the cost function $c(s_{t}, a_{t})$ of BC is a differentiable training loss function, with discount factor $\gamma_{t}\equiv1$, and the dynamics of BC depends only on the data distribution, independent from the current $(s_{t}, a_{t})$ pair.

Training a model in BC paradigm is identical to fitting a supervised regressor. For instance, one can fit a Neurual Network (NN) regressor with the cost function $c(\cdot,\cdot)$ set to L2 loss:
\begin{equation}
\label{eq:dnn}
a_{t} = f_{NN}\left( \left. s_{t} \right\vert \theta_{NN}\right), 
\end{equation}
where $s_{t}$ is the state of an expert agent, including its local visibility (e.g., a range map, a velocity map) from the center point of this agent, as well as a local guidance velocity and a global guidance velocity -- see details on state representation in the evaluation part. $\theta_{NN}$ is the model parameter. Such NN based model can also represent many traditional regressors including support vector regressor, random forest, etc.

As mentioned earlier, in crowd simulation agents are goal directed. To arrive the final destination, it is critical for the state $s_{t}$ of an agent to contain not only the local observation about where neighboring agents/obstacles exist and what the relative velocities of neighboring agents/obstacles are wrt the agent, but also a local guidance direction (or local guidance velocity) that leads the agent to its own nearest sub-goal location. Such local guidance velocity is agent-specific and dependent on the current location of the agent. In addition, due to the existence of environmental obstacles, the local guidance velocity may not coincide with the global guidance velocity that directly points to the final destination of the agent.

The local guidance velocity can be either learned from experience (e.g., from expert trajectories) or planned by an external planner provided with the environment configuration and the initial/destination positions of an agent. When the movement of expert agents forms a flow pattern, indicating that two nearby agents have similar trajectories, the flow can be learned with a Gaussian Process (GP). With the learned GP, when an imitator is generalized to that environment, the prediction of the GP could provide the local guidance velocity for the imitator in $s_{t}$:
\begin{equation}
a_{t}^{\text{local}} \sim GP\left(\, \cdot \, \vert (x,y,t), \textbf{X}_{\text{train}}, \theta_{GP}\right), 
\end{equation}
where $(x,y,t)^{T}\in R^{3}$ is the spatial-temporal location of the imitator, $\textbf{X}_{\text{train}}$ is the training data, $\theta_{GP}$ is the hyper-parameter, and $a_{t}^{\text{local}}$ is the local guidance velocity at the current spatial-temporal location.

On the other hand, when the movement of expert agents does not form a flow pattern, one may use a path-planning algorithm to provide the local guidance velocity in $s_{t}$.

\section{Reinforcement Learning Agents}
\label{sec5}
Reinforcement learning first estimates $c^{*}$ from expert trajectories, then estimates the optimal policy $\pi^{*}$ to approximate the underlying but unknown expert policy $\pi_{E}$. One approach to recover $c^{*}$ is the maximum causal entropy IRL~\cite{ziebart2008maximum}:
\begin{align}
\argmax_{c \in \mathcal{C}} \min_{\pi \in \Pi} -H(\pi) + \mathbb{E}_{\pi}\left[ c(s, a) \right] - \mathbb{E}_{\pi_{E}}\left[ c(s, a) \right], 
\end{align}
where $H(\pi) \triangleq \mathbb{E}_{\pi}\left[ -\log \pi(a|s) \right]$, $\mathcal{C}$ is the family of cost functions, $\Pi$ is the family of policy functions, and $\pi_{E}$ denotes the expert policy that generates the expert trajectories. Here $c^{*}$ minimizes the expected cost of expert trajectories while maximizes the cost of the policy trajectories. If such $c^{*}$ is obtained, the optimal policy $\pi^*$ satisfies
\begin{align}
\argmin_{\pi \in \Pi} -H(\pi) + \mathbb{E}_{\pi}\left[ c(s, a) \right], 
\end{align}
and can be estimated in a regularized RL procedure.

The two-step IRL-RL are complex. Recently, ~\cite{ho2016generative} have proposed GAIL, in which they have shown that the two-step IRL-RL is identical to a one-step occupancy matching procedure.

To induce GAIL paradigm, they first add a closed, proper convex cost function regularizer $\psi : \mathbb{R}^{\mathcal{S} \times \mathcal{A}} \rightarrow \mathbb{R}$ to alleviate the overfitting issue stemming from the finite dataset size. With this regularizer, the IRL objective is given by
\small
\begin{align}
\argmax_{c \in \mathcal{C}} & 
	- \psi(c) + \left( \min_{\pi \in \Pi} -H(\pi) + \mathbb{E}_{\pi}\left[ c(s, a) \right] \right) - \mathbb{E}_{\pi_{E}}\left[ c(s, a) \right]. 
\end{align}
\normalsize

On the other hand, they define an occupancy measure $\rho_{\pi} : \mathcal{S} \times \mathcal{A} \rightarrow \mathbb{R}$ of a stochastic policy $\pi$ as
$\label{eq:occupancy_measure_def}
\rho_{\pi}(s, a) = \pi(a|s) \sum_{t} \gamma_{t} P(s_{t} = s | \pi)$. 
$\rho_{\pi}$ describes the distribution of $(s, a)$ pairs that an agent encounters when navigation with policy $\pi$. (policy is stochastic in training but deterministic in testing to trade off exploitation for exploration).

With this definition, they show that RL and IRL solve the primal and the dual problems of occupancy measure matching, with optimal solutions forming a saddle point. That means any two-step IRL-RL is equivalent to the following one-step formulation:
\begin{align}
\label{eq:one_step_via_occupancy_measure}
\pi^{*} = \argmin_{\pi \in \Pi} \psi^{*}(\rho_{\pi} - \rho_{\pi_{E}}) -\lambda H(\pi)
\end{align}
where $\psi^{*}$ (the convex conjugate of function $\psi$) is the convex function measuring the deviation of $\rho_{\pi}$ from $\rho_{\pi_{E}}$. This suggests that finding $\pi^{*}$ to approach $\pi_{E}$ can be transformed to matching the occupancy measure between $\rho_{\pi}$ and $\rho_{\pi_{E}}$. Here $\lambda$ is an introduced regularization parameter to control the entropy term.

They further show that there exists a specific $\psi$:
\begin{align}
\psi_{GA}(c) &\triangleq
	\begin{cases}
    \mathbb{E}_{\pi_{E}} \left[ g(c(s,a)) \right] &\quad \text{if } c < 0 \\
    +\infty &\quad \text{otherwise} 
    \end{cases}
\end{align}
where $g(x) = -x -\log (1 - e^{x})$ if $x < 0$, otherwise $g(x)=+\infty$,
such that $\psi_{GA}(\rho_{\pi} - \rho_{\pi_{E}})$ can be represented as:
\small
\begin{align}
\psi_{GA}^{*}(\rho_{\pi}-\rho_{\pi_{E}}) = \max_{D} \mathbb{E}_{\pi} \left[ \text{log}(D(s,a)) \right] + \mathbb{E}_{\pi_{E}}\left[ \text{log}(1-D(s,a)) \right]  \nonumber
\end{align}
\normalsize
where $D: \mathcal{S} \times \mathcal{A} \rightarrow (0,1)$, which is employed to predict the probability that a given state-action pair comes from $\pi$ rather than $\pi_{E}$, with the relation $c(s,a) = \text{log}D(s,a)$. 

In that case, the one-step formulation in Eq.~\ref{eq:one_step_via_occupancy_measure} is reduced to an adversarial form:
\small
\begin{align}
\label{eq:gail_final}
\min_{\pi} \max_{D} &\, \mathbb{E}_{\pi} \left[ \log(D(s,a)) \right] + \mathbb{E}_{\pi_{E}} \left[ \log(1 - D(s,a)) \right] - \lambda H(\pi).
\end{align}
\normalsize

Therefore, the final objective given by Eq.~\ref{eq:gail_final} can be optimized adversarially with gradient descend and policy optimization (e.g., trust region policy optimization ~\cite{schulman2015trust}). Eventually, both cost function and policy function can be obtained simultaneously, capable of representing a general two-step IRL-RL models.

\section{Data Domains}
\label{sec6}

We identify three scenario domains in this work: exocentric standard scenarios (\texttt{X}), egocentric representative scenarios (\texttt{G}), and egocentric random scenarios (\texttt{R}). In all domains, an agent is represented as a circle with the radius of 0.5 meters.

\begin{figure*}[htb]
\small
\begin{tabular}{c c c c c c}
 \includegraphics[width=0.16\linewidth]{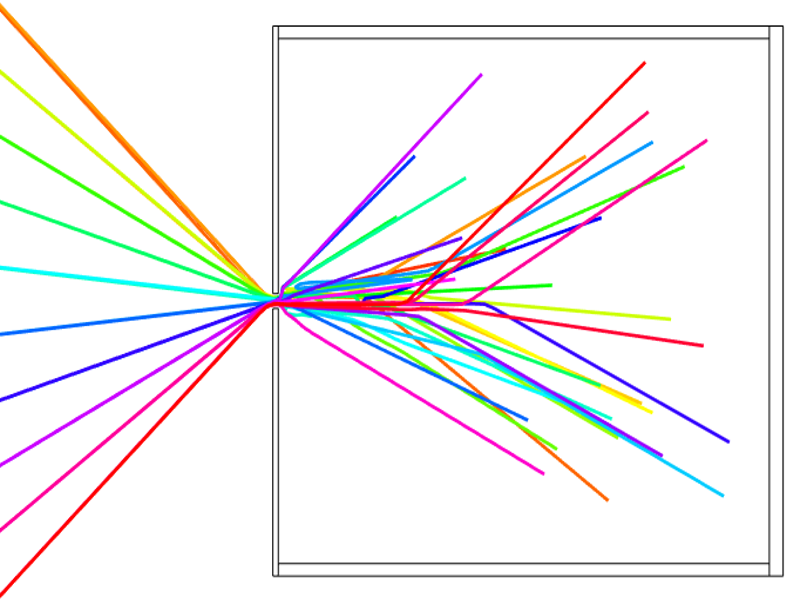}
~~~~
 & \includegraphics[width=0.12\linewidth]{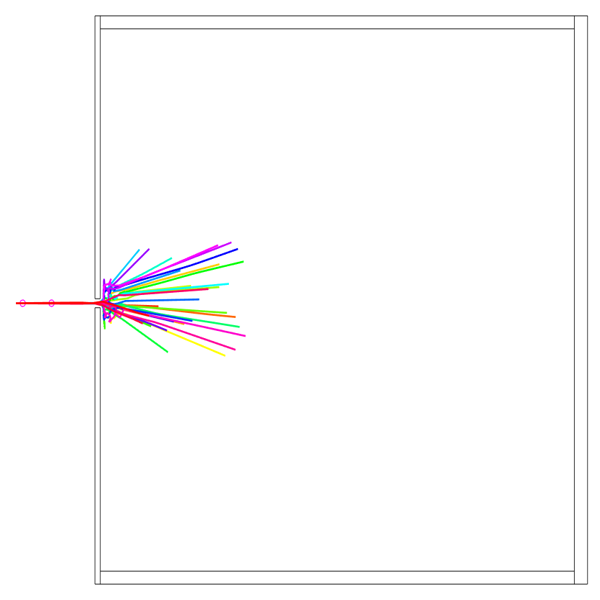}
~~~~
& \includegraphics[width=0.12\linewidth]{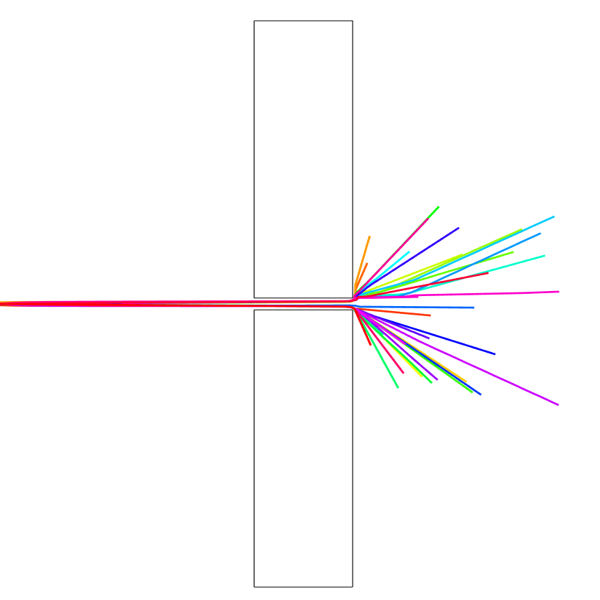}
~~~~
 & \includegraphics[width=0.12\linewidth]{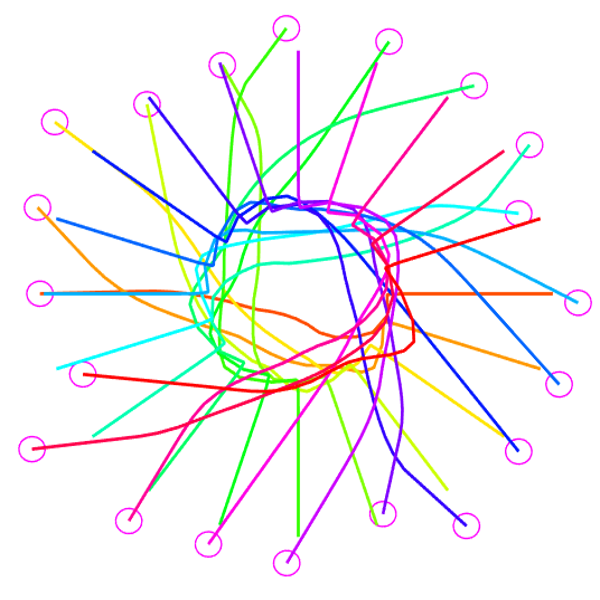}
~~~~
 & \includegraphics[width=0.12\linewidth]{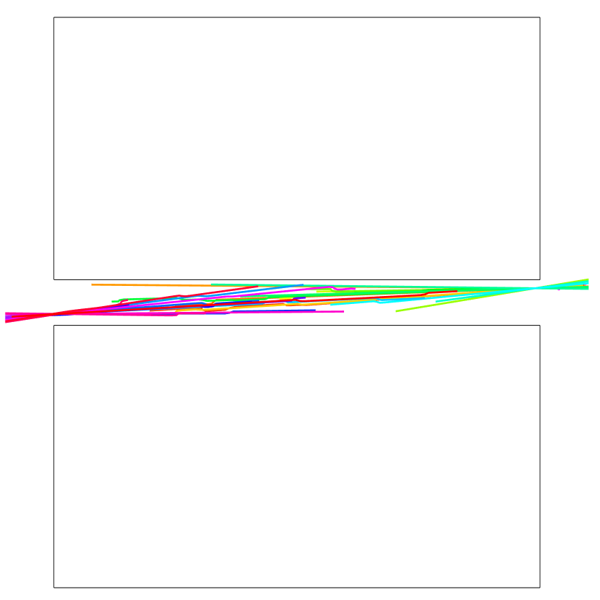}
~~~~
 & \includegraphics[width=0.12\linewidth]{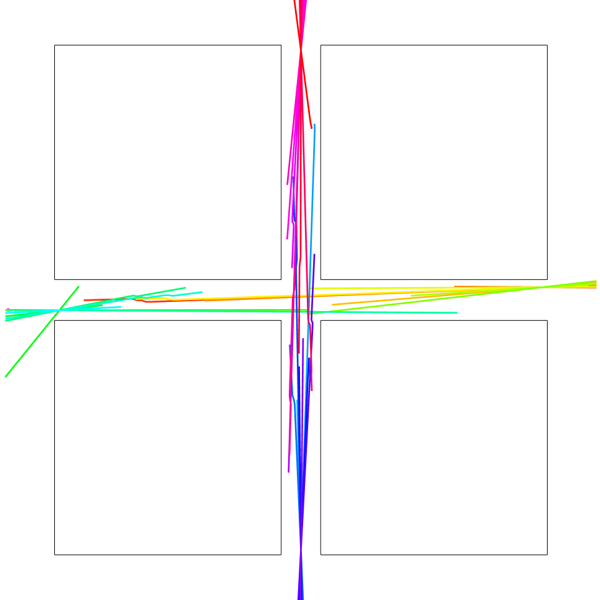} \\ 
 (1) & (2) &\hspace{-0.4cm} (3) & \hspace{-0.4cm} (4) & \hspace{-0.3cm} (5) & (6) \\ 
 \end{tabular}
\caption{Exocentric standard scenarios (\texttt{X}). (1) Evacuation 1, (2) Evacuation 2, (3) Bottleneck squeeze, (4) Concentric circles, (5) Hallway two-way, (6) Hallway four-way.}
\label{fig:6scenario-dataset}
\end{figure*}

\begin{figure}[htb]
 \begin{subfigure}{0.23\textwidth}
 \centering
 \includegraphics[width = \textwidth]{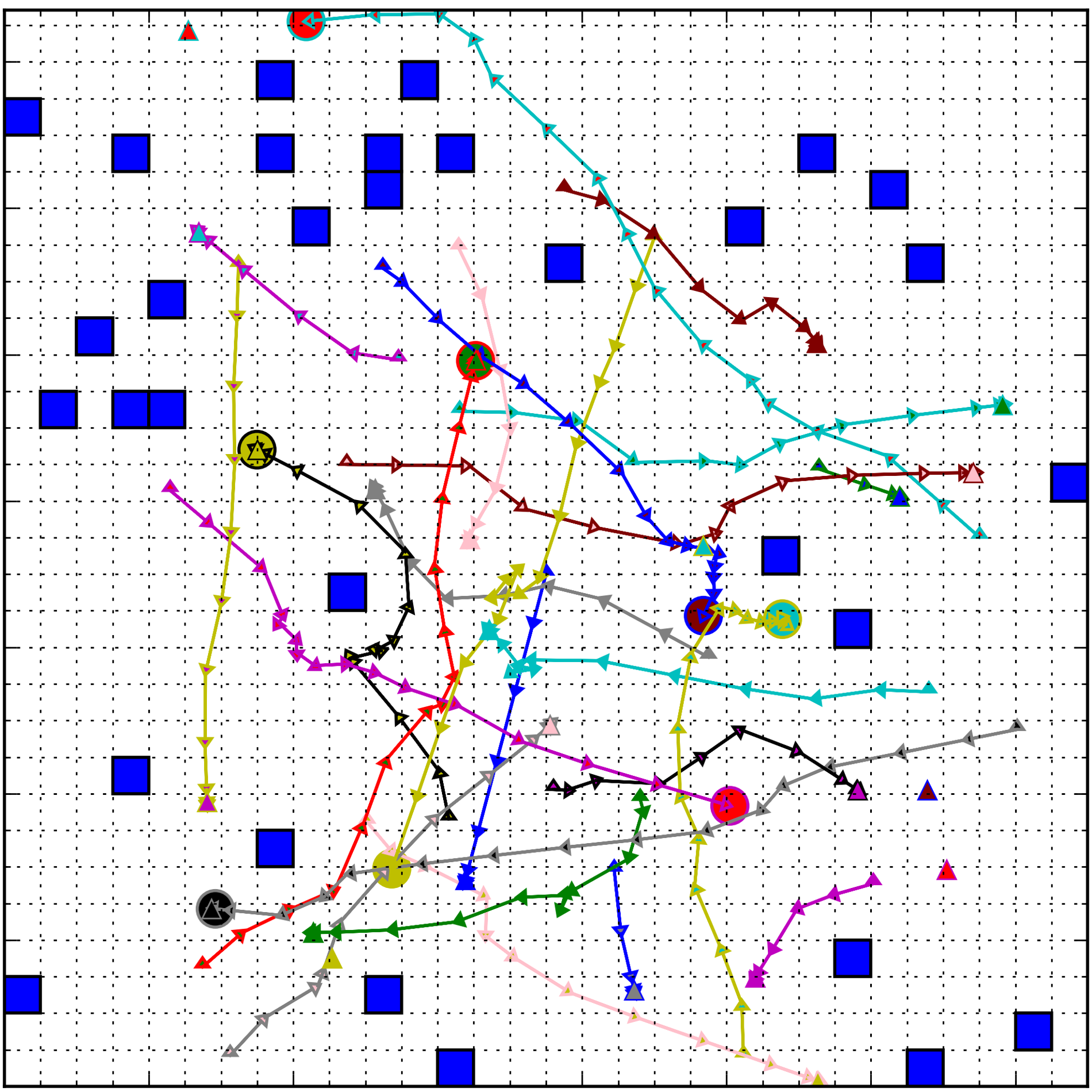}
 \end{subfigure}
 \begin{subfigure}{0.23\textwidth}
 \centering
 \includegraphics[width = \textwidth]{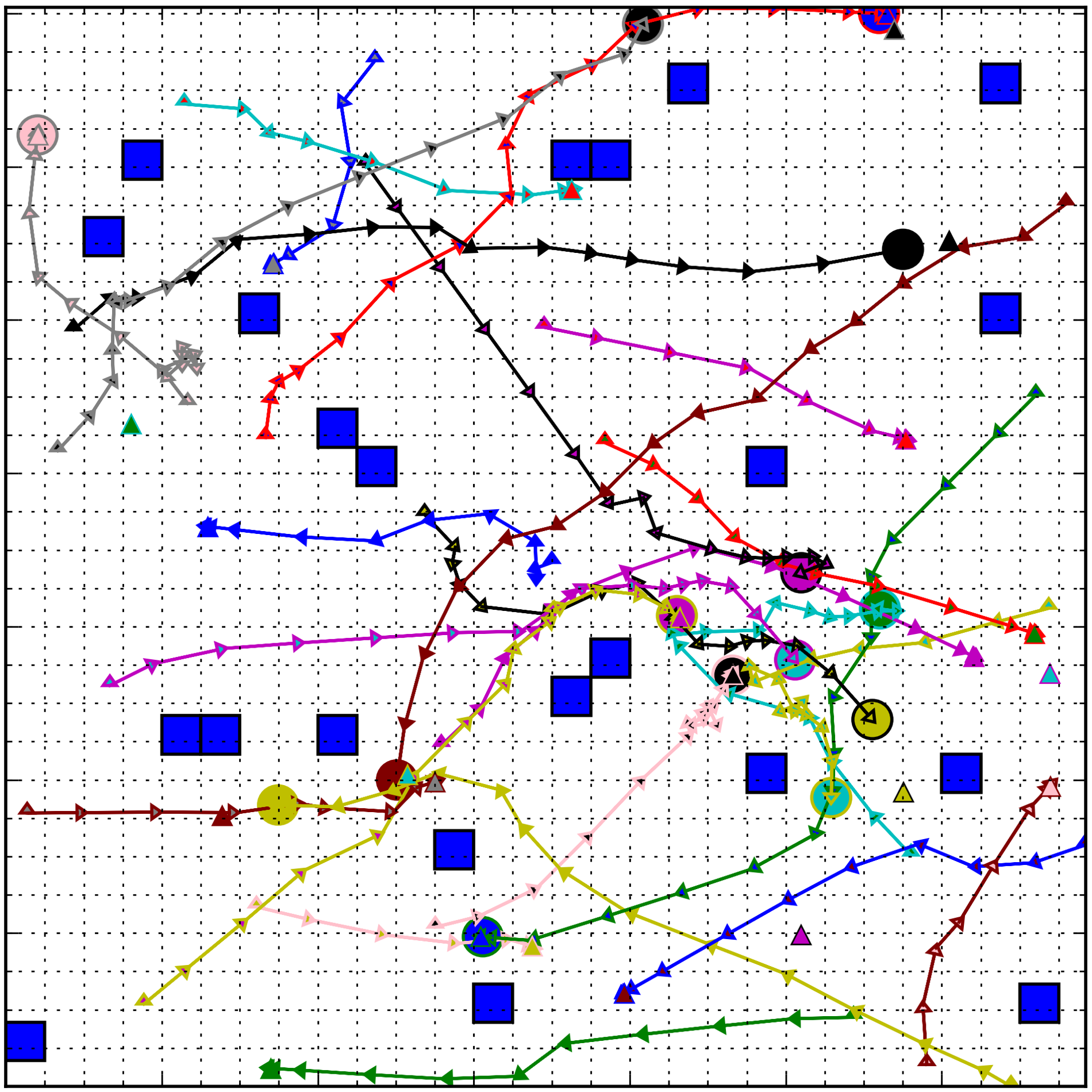}
 \end{subfigure}
\caption{Example scenarios from egocentric representative (\texttt{G}) domain, shown with expert trajectories. Each agent (denoted as a circle) aims to reach its destination (a triangle of the same color) while avoiding other agents and obstacles.}
\label{fig:generic-dataset}
\end{figure}

\subsection{Exocentric Standard Scenarios (\texttt{X})}
This domain provides a few but complex and crowded scenarios, including six environment benchmarks used to evaluate computational models of crowd movement ~\cite{singh2009steerbench,yoon2016filling,qiao2018role}. The six scenarios
(with variation in agent density and initial/destination positions) include:

\setlength{\leftmargini}{0.5cm}
\begin{enumerate}
\item \textbf{Evacuation 1.} Many agents must evacuate a room, with only one small doorway of width 2.4 m. Agents are heading toward distinct target locations outside the room.

\item \textbf{Evacuation 2.} Similar to Evacuation 1 but the doorway width is narrowed to 1.4 m. Also agents are heading toward the same target location outside of the room.

\item \textbf{Bottleneck squeeze.} All agents begin on one side of the area, and enter and traverse a hallway to reach the target. 

\item \textbf{Concentric circles.} Agents are symmetrically placed along a circle and aim to reach antipodal positions.

\item \textbf{Hallway two-way.} Many agents traveling in either direction through a hallway. Agents are expected to form lanes. 

\item \textbf{Hallway four-way.} Many agents arriving from and traveling to any of the four cardinal directions. 
\end{enumerate} 
Illustrations of the six scenarios are shown in Fig.~\ref{fig:6scenario-dataset}.

\subsection{Egocentric Representative Scenarios (\texttt{G})}
Exocentric standard scenarios provide challenging crowd tasks, but may not be able to sufficiently provide a \emph{representative space} of challenging local interactions individuals encounter in crowd. Egocentric random scenarios provide random samples of state-action pairs, but these samples can not form complete trajectories, and there are no agent-obstacle interactions.

In an effort to produce a data domain with a large number of sampling of small-scale, but general inter-agent and agent-obstacle interactions that individuals encounter, we refer to ~\cite{kapadia2011scenario}, which characterizes the representative space of scenarios observed in crowds, and a sampling strategy to generate a finite set of scenarios with sufficient coverage. Specifically, a considerable amount of simulation scenarios are uniformly sampled from this scenario space for both training and testing ($4000$ for training, $100$ for testing).  Each scenario contains randomly distributed obstacles and randomly assigned initial/destination positions of agents, with expert driven by the social force model. Fig.~\ref{fig:generic-dataset} illustrates two samples in this domain.

\subsection{Egocentric Random Scenarios (\texttt{R})}
The randomly generated scenarios proposed in~\cite{long2017deep} constitute the second domain, where a sufficient number of samples are collected by uniformly and independently sampling over the state space. The positions of neighboring agents, previous velocities of neighboring agents and the preferred velocity of a reference agent are randomly set to construct a particular state for the reference agent at a step, while the expert decision of the reference agent at this step is queried from ORCA~\cite{van2011reciprocal} given the same state. This produces many discrete and independent snapshots for immediate responses of an expert to inter-agent interactions. Note that in any sample of this domain, there are no obstacles, thus no agent-obstacle interactions involved.

\subsection{Summary of Three Data Domains}
In the following sections, abbreviations \texttt{X}, \texttt{G}, and \texttt{R} are used to denote the domain of exocentric standard, egocentric random and egocentric representative scenarios respectively. Tab.~\ref{tab:summary_of_data_domains} summarizes the characteristics of each domain.

\begin{table}[t] 
\centering
\caption{characteristics of three data domains}
\vspace{-2mm}
\begin{tabular}{|c|c|}
\hline
\texttt{X} & \begin{tabular}[c]{@{}c@{}}A few challenging obstacle configurations;\\with inter-agent interactions\end{tabular} \\ \hline
\texttt{G} & \begin{tabular}[c]{@{}c@{}}Diverse inter-agent and agent-obstacle interactions;\\many test scenarios\end{tabular}   \\ \hline
\texttt{R} & \begin{tabular}[c]{@{}c@{}}Numerous diverse snapshots on inter-agent\\interactions; no complete trajectories;  no obstacles\end{tabular}  \\ \hline
\end{tabular}
\label{tab:summary_of_data_domains}
\end{table}

\section{Evaluating Scenario Generalization Capability} \label{sec:experiments}
Bidirectional experiments are conducted: models trained on egocentric representative (\texttt{G}) and egocentric random (\texttt{R}) are tested on exocentric standard scenarios (\texttt{X}); models trained on exocentric standard (\texttt{X}) and egocentric random (\texttt{R}) are tested on egocentric representative scenarios (\texttt{G}).

\subsection{Trained Models}
Given the two training paradigms and three data domains, five training paradigm -- training domain combinations are studied:
\setlength{\leftmargini}{0.5cm}
\begin{enumerate}

\item \texttt{BCA-X}: BC agents trained on \texttt{X}

\item \texttt{BCA-G}: BC agents trained on \texttt{G}

\item \texttt{BCA-R}: BC agents trained on \texttt{R}

\item \texttt{RLA-X}: RL agents trained on \texttt{X}

\item \texttt{RLA-G}: RL agents trained on \texttt{G}

\end{enumerate}
RL agents are not trained on egocentric random scenarios as RL require complete trajectories, not independent state-action pairs.

\subsection{State Representation}
Similar to~\cite{qiao2018role}, we simulate that each agent observes the world around it using a collection of local measurements.  The first local measurement is a range map, a measure of radial distances from the center of the agent to the surface of the environment (including surfaces of neighboring agents and surfaces of obstacles), typically at a resolution of one degree over 360 degrees. We also simulate that an agent can detect the relative movements of neighboring agents and obstacles, perceiving a radial velocity map. In addition, an agent receives local and global guidance velocities. The local guidance velocity is provided by an external source (either GP or A-star), which is capable of sensing obstacles in the environment but lacks knowledge of the existence of other moving agents, thus guiding the agent's movement independent of other agents, like a GPS.  The global signal provides an overall heading direction towards the final destination position, much like a compass.

Following~\cite{qiao2018role,long2017deep}, GP provides the local guidance velocity in exocentric standard scenarios, while the sampled preferred velocity acts as the local guidance in egocentric random scenarios. However, in egocentric representative scenarios, the movement of agents does not form a flow pattern. Therefore, we use A-star to plan a route for each agent from its initial to its destination position. Influenced by neighboring moving agents, an agent does not follow strictly with its A-star way points. Instead, at each step it aims at its furthest A-star way point it sees without visual occlusion as the current local goal.

\subsection{Main Training Configuration}
For the size of the training data, the amount of state-action pairs for training in three domains are nearly the same, about 1.6M.

All \texttt{BCA-X}, \texttt{BCA-G}, \texttt{BCA-R} adopt a six-layer fully connected network, with each layer containing 100 neurons. They are trained by RMSprop~\cite{tieleman2012divide} with L2 loss and learning rate 0.0001.

For training reinforcement learning (RL) agent, both policy and reward functions adopt the same architecture as \texttt{BCA-X}, \texttt{BCA-G}, \texttt{BCA-R} to ensure that all policies share the same model complexity. The policy learning rate for RL agent is set to 0.01. During sampling model trajectories in the training phase, a zero-mean Gaussian random noise with standard deviation 0.5 is added to the output trading off for exploration. The policy entropy regularizer $\lambda$ is set to be 0. The network is trained at 10K iterations for exocentric standard scenarios and 6K iterations for egocentric representative scenarios.

\subsection{Metrics} 
The five models are evaluated on three metrics, following ~\cite{qiao2018role}. All metrics are the lower, the better.

\begin{enumerate} 
\item\textbf{DTW metric}: 
Dynamic Time Warping distance ~\cite{salvador2007toward} measures the spatial deviation of a model trajectory from an expert trajectory averaged over agents. To eliminate the influence of different number of steps in model trajectories, a min-match version of DTW is adopted, by registering each of the nodes (positions) of a model trajectory to its closest node of the corresponding expert trajectory using dynamic programming, and accumulating the minimal distance of registered pairs of each node along the expert trajectory.

\item\textbf{AA metric}: 
AA stands for agent-agent collisions, the total number of collisions for all pairs of agents accumulated over all steps. During one-step movement, a collision between one pair of agents occurs if their distance is less than the sum of their radii at any real-valued time point within that time duration, which could be verified by solving a distance-related quadratic equation.

\item\textbf{AO metric}:
AO denotes agent-obstacle collisions, the total number of collisions between all pairs of an agent and an edge of an obstacle during a simulation, also accumulated over timesteps. An agent-obstacle collision can be detected based on (1) the intersection of two line segments (one for an edge of an obstacle, the other for the trace of an agent's center during a one-step movement) and (2) the distance between a point (the center of an agent) and a line segment (an edge of an obstacle).
\end{enumerate}

Note that within one step if an agent collides with more than one edge of an obstacle, only one AO collision is counted. Two agents keep overlapping or an agent moving within an obstacle is only counted once for the first contacting of their edges until they depart from each other. Also for simplicity, if an agent-agent or agent-obstacle collision occurs, it does not change the velocity of involved agents within that temporal duration. 

\begin{figure*}[t]
 \centering
 \includegraphics[width=13.0cm]{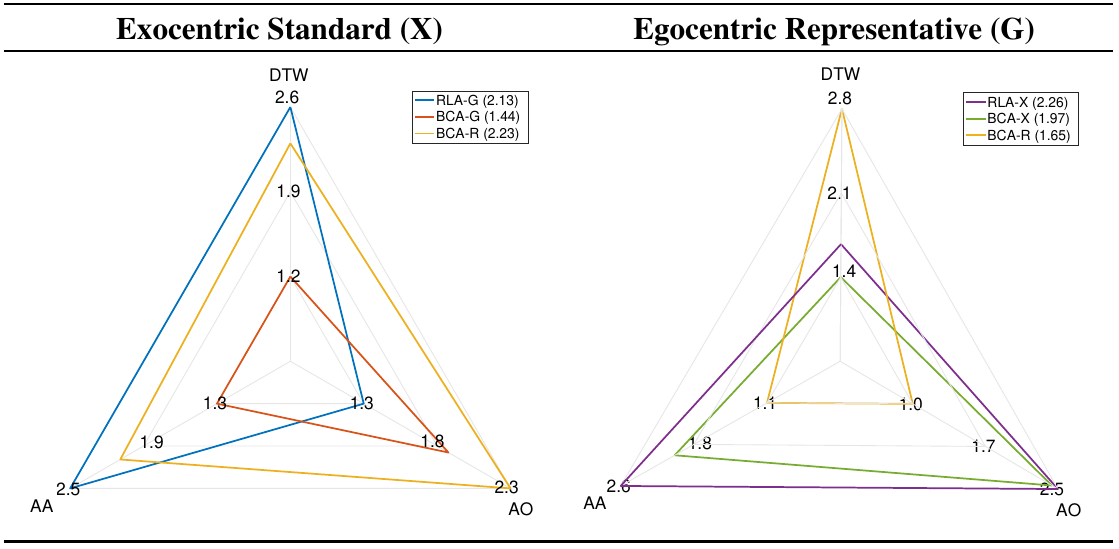}
 \caption{Rankings of models over test domains. Each color represents a model, and each axis indicates averaged rankings over test scenarios for a metric. For all three metrics, the smaller the better. Number in legend denote averaged ranking over three metrics.}
\label{fig:ranking_comparison_float}
\end{figure*}

\begin{table*}[t] 
  \caption{Comparison of \texttt{RLA-G} (blue), \texttt{BCA-G} (red) and \texttt{BCA-R} (yellow), using the three metrics with different agent densities increasing from left to right. Each axis in a plot denotes one type of the six scenarios: A, B, C, D, E and F denotes evacuation 1, evacuation 2, bottleneck squeeze, concentric circles, hallway two-way, and hallway four-way, respectively.  Line thickness in plots indicates each metric's standard deviation. Polygons closer to the origin imply a better (lower) metric value. } \label{tab:spider_metrics_on_F}
  \begin{tabular}{p{0.45cm}p{3.0cm}p{3.0cm}p{3.0cm}p{3.0cm}p{3.0cm}} \hline 
      \multicolumn{1}{>{\centering\arraybackslash}m{0.3cm}}{Density} 
      & \multicolumn{1}{>{\centering\arraybackslash}m{3.0cm}}{10} 
      & \multicolumn{1}{>{\centering\arraybackslash}m{3.0cm}}{20} 
      & \multicolumn{1}{>{\centering\arraybackslash}m{3.0cm}}{30}
      & \multicolumn{1}{>{\centering\arraybackslash}m{3.0cm}}{40} 
      & \multicolumn{1}{>{\centering\arraybackslash}m{3.0cm}}{50}\\ \hline
      
      \textbf{DTW} 
      & \parbox[c]{1em}{\includegraphics[width=3.2cm]{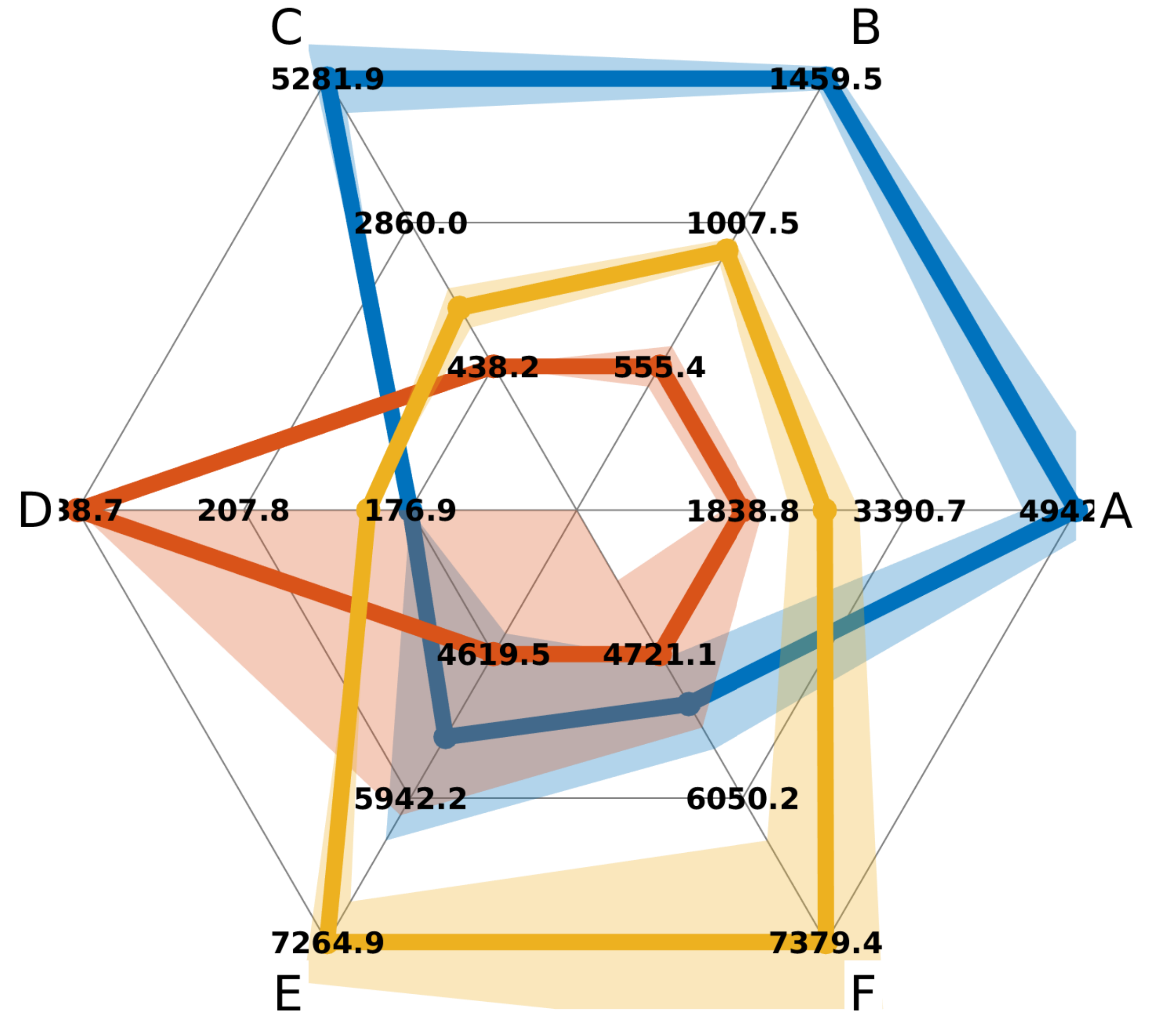}}    &
      \parbox[c]{1em}{\includegraphics[width=3.0cm]{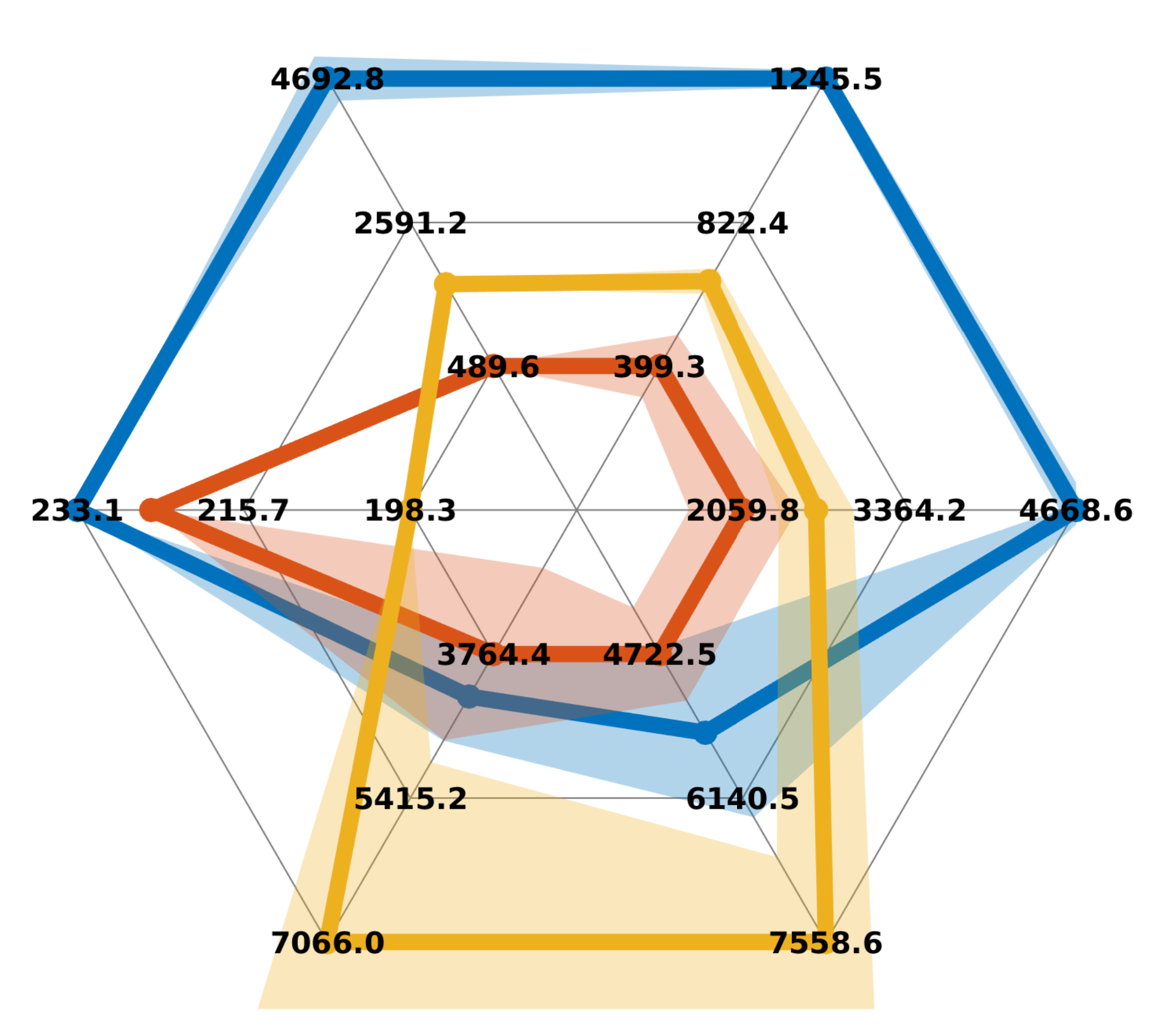}}
      & \parbox[c]{1em}{\includegraphics[width=3.0cm]{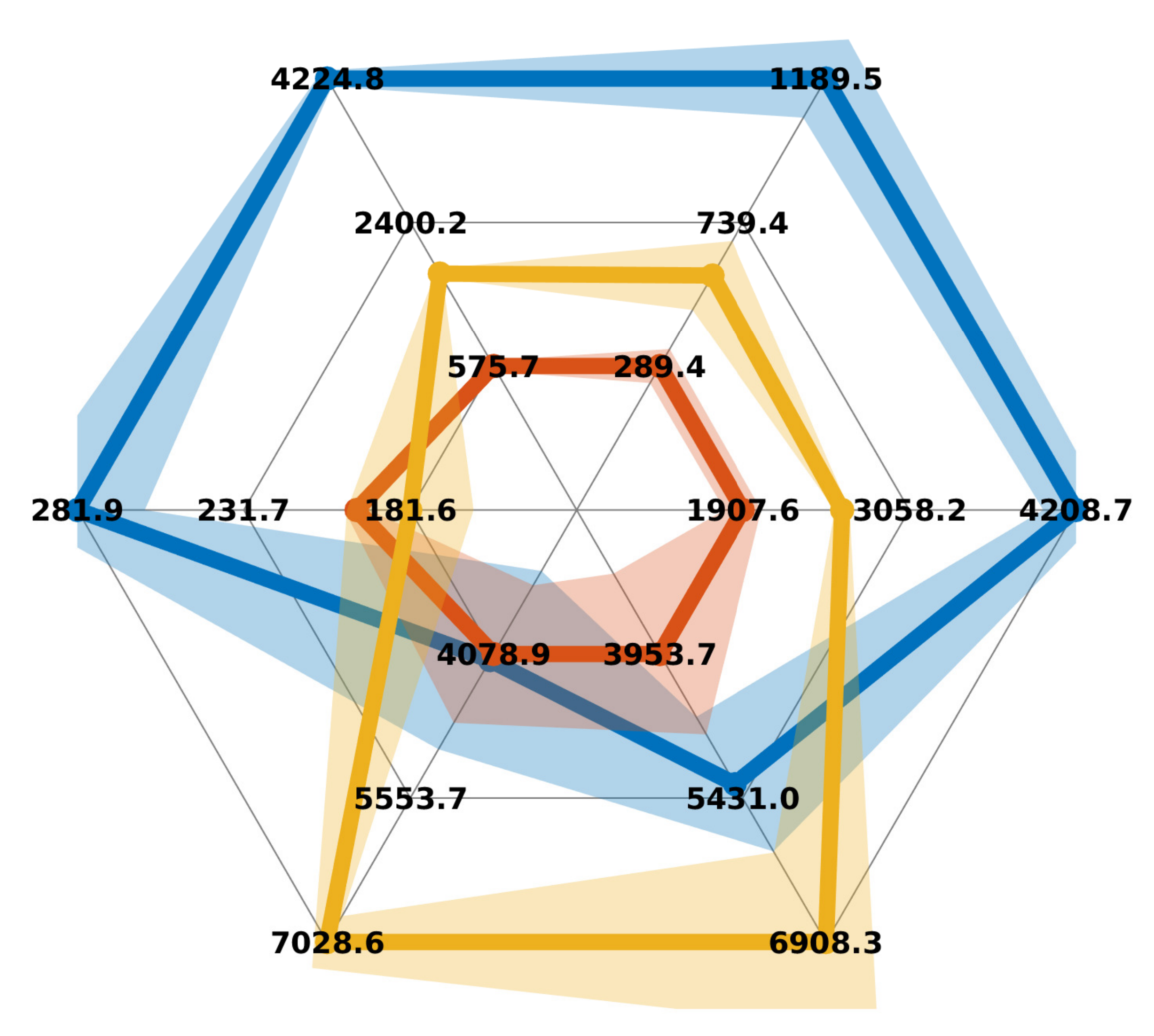}} & 
      \parbox[c]{1em}{\includegraphics[width=3.0cm]{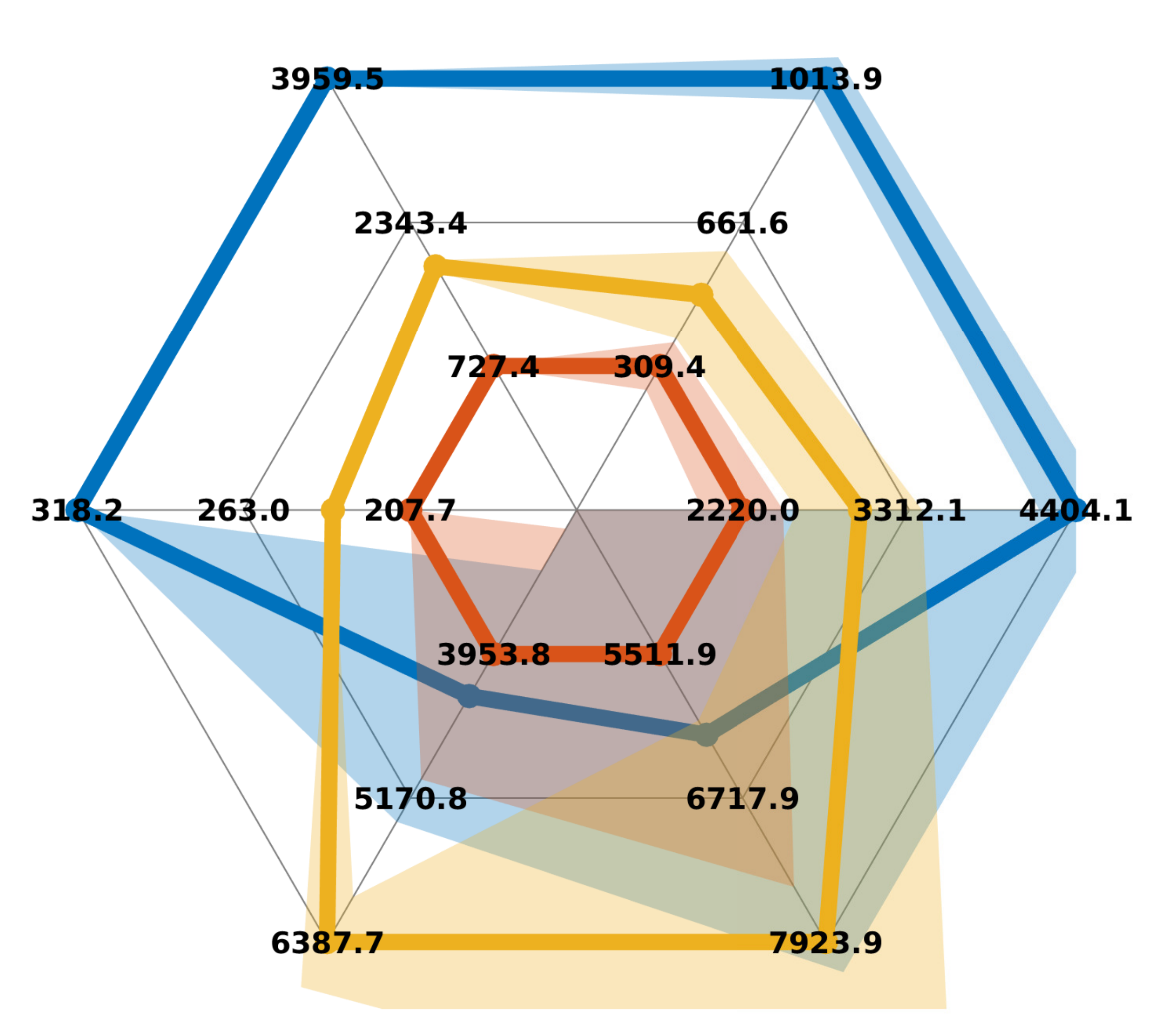}}
      & \parbox[c]{1em}{\includegraphics[width=3.0cm]{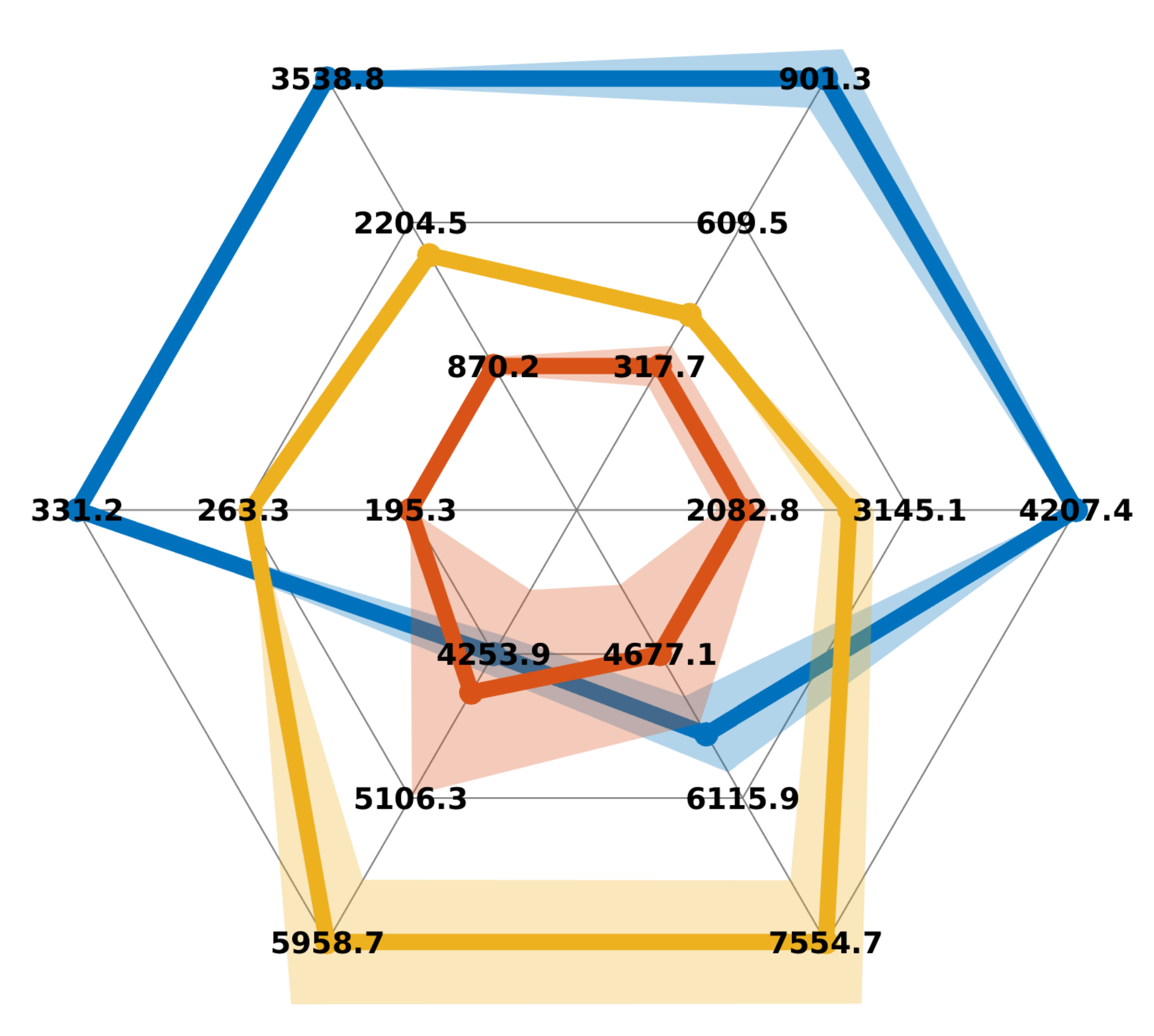}} \\ \hline
      
      \textbf{AA}  
      & \parbox[c]{1em}{\includegraphics[width=3.0cm]{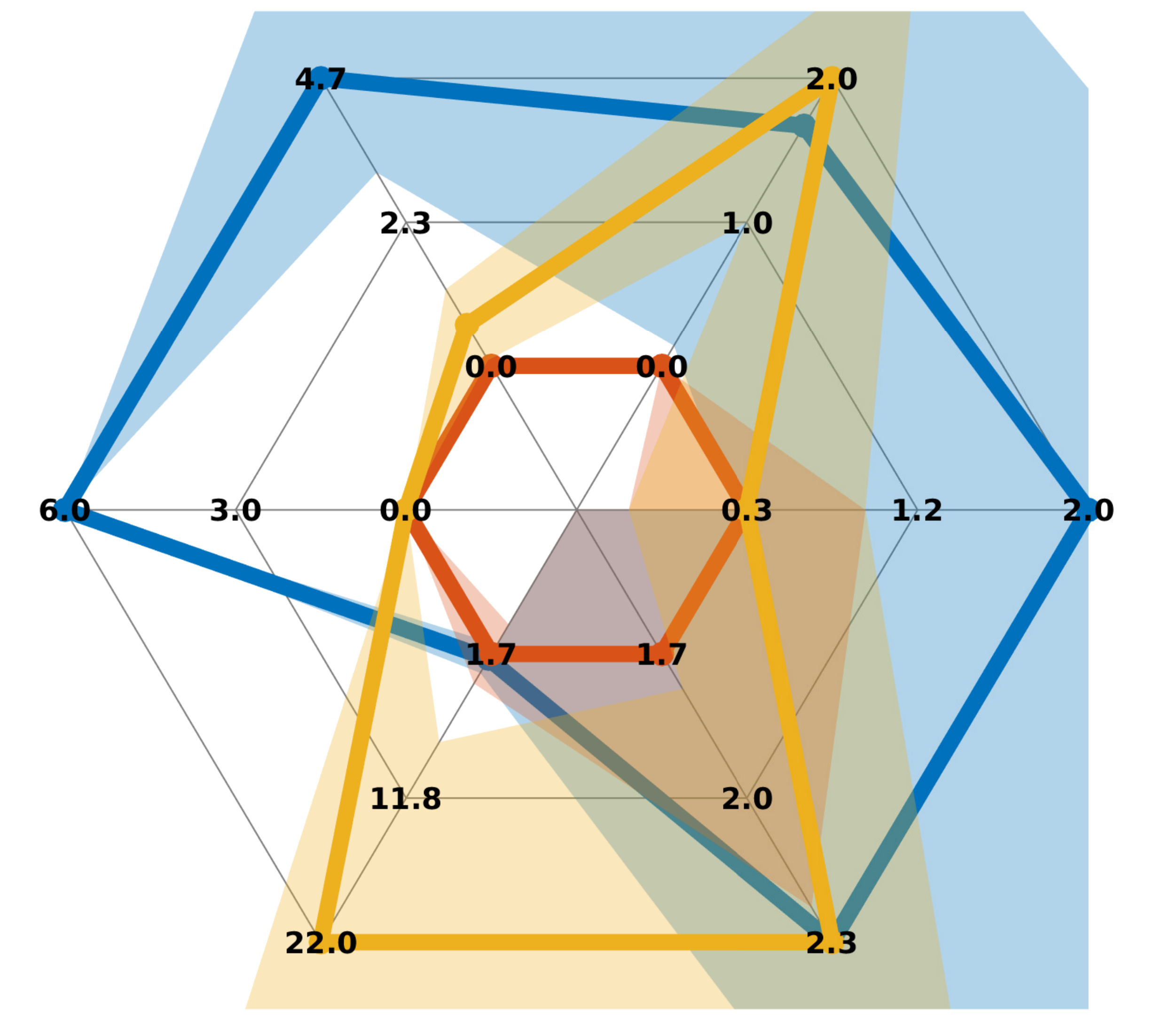}}     & 
      \parbox[c]{1em}{\includegraphics[width=3.0cm]{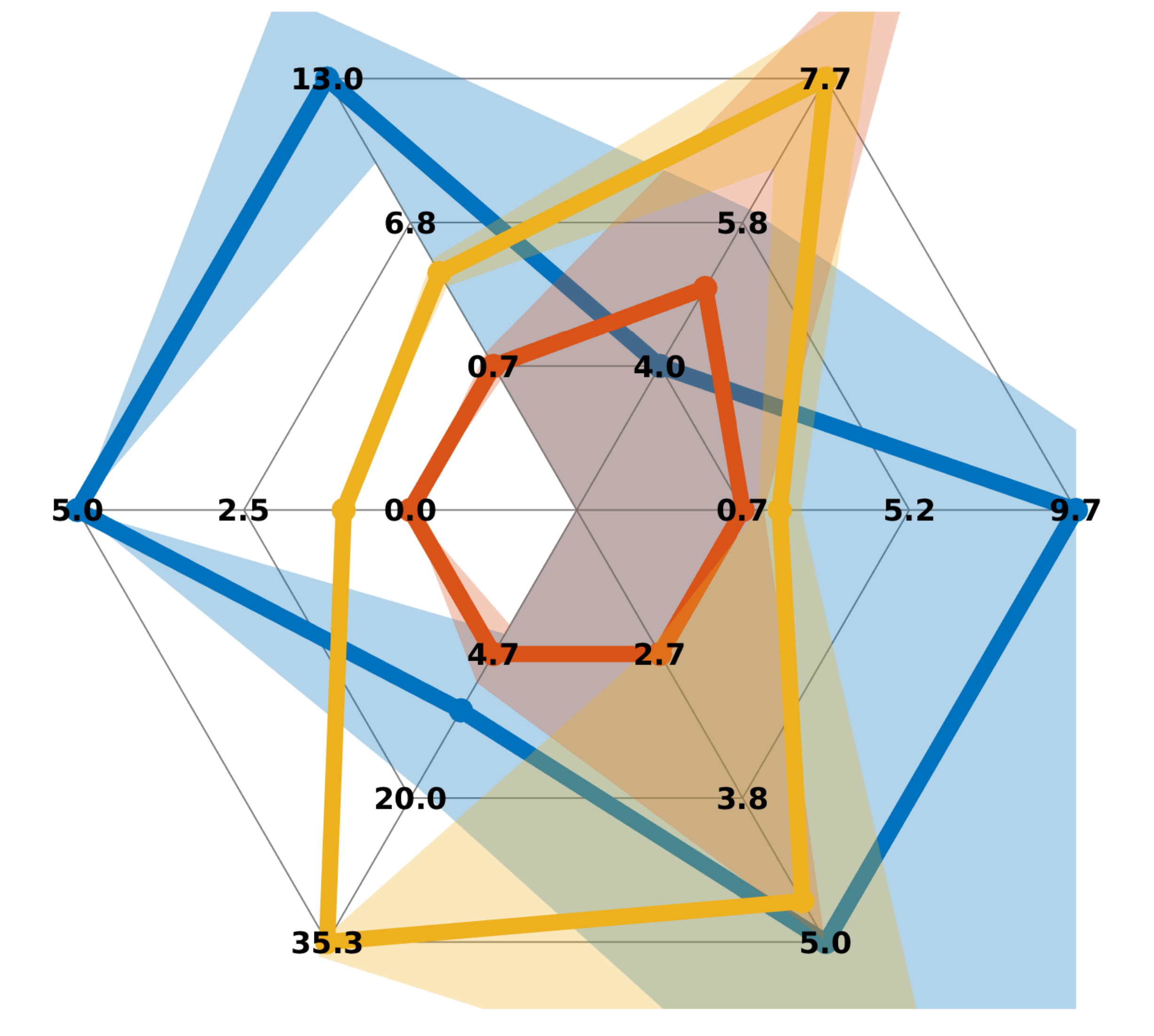}}  
      & \parbox[c]{1em}{\includegraphics[width=3.0cm]{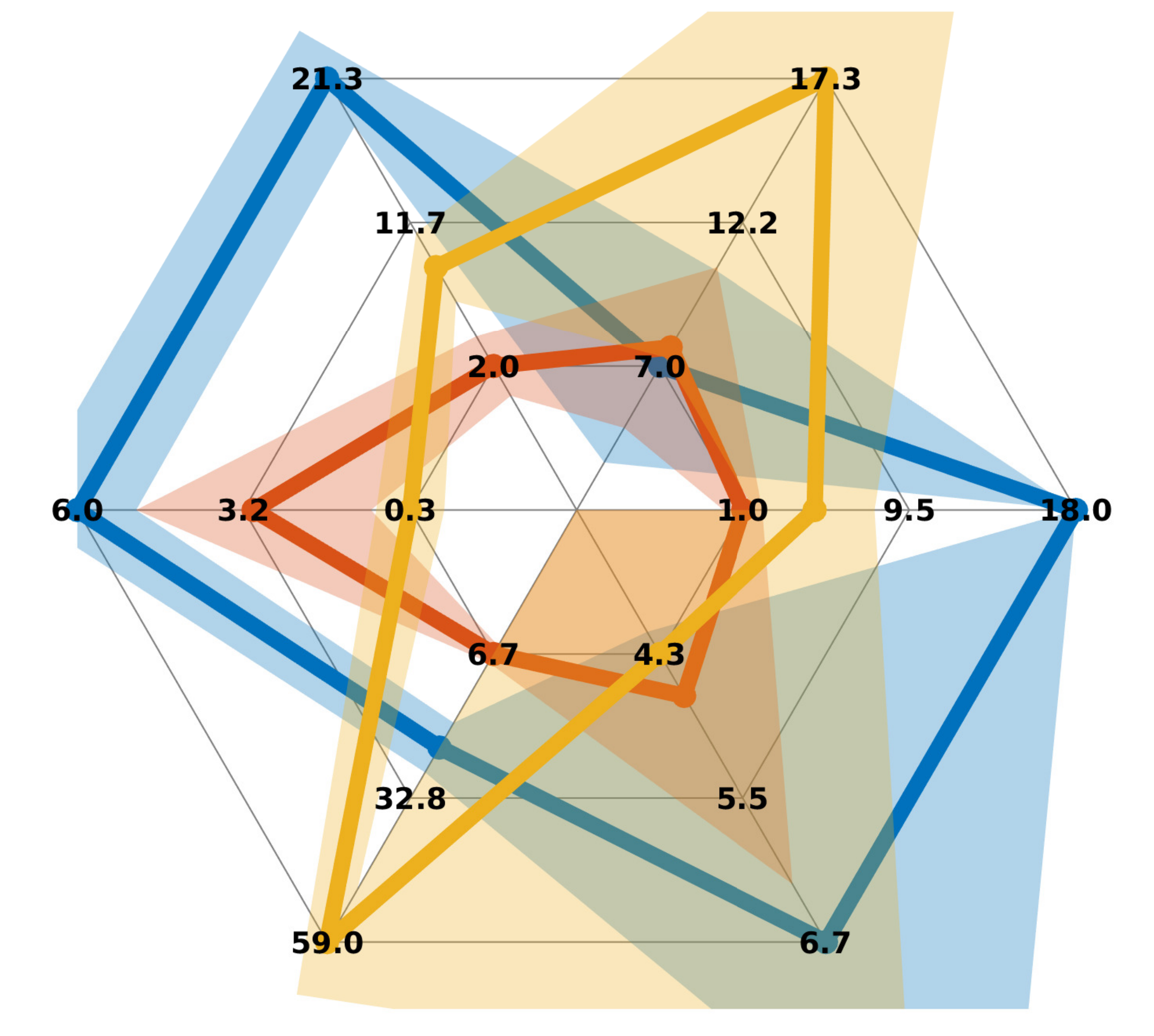}}  & 
      \parbox[c]{1em}{\includegraphics[width=3.0cm]{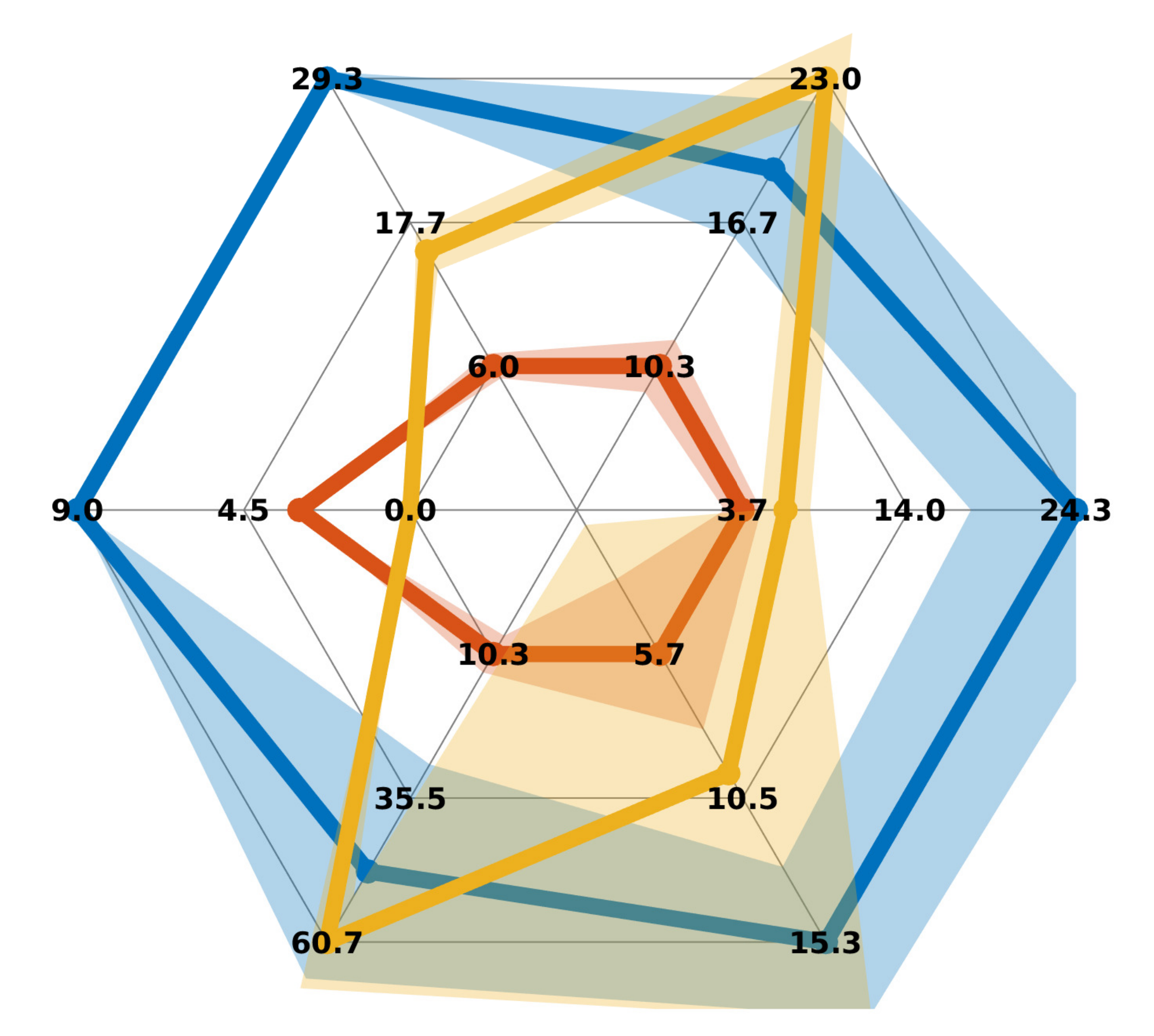}} 
      & \parbox[c]{1em}{\includegraphics[width=3.0cm]{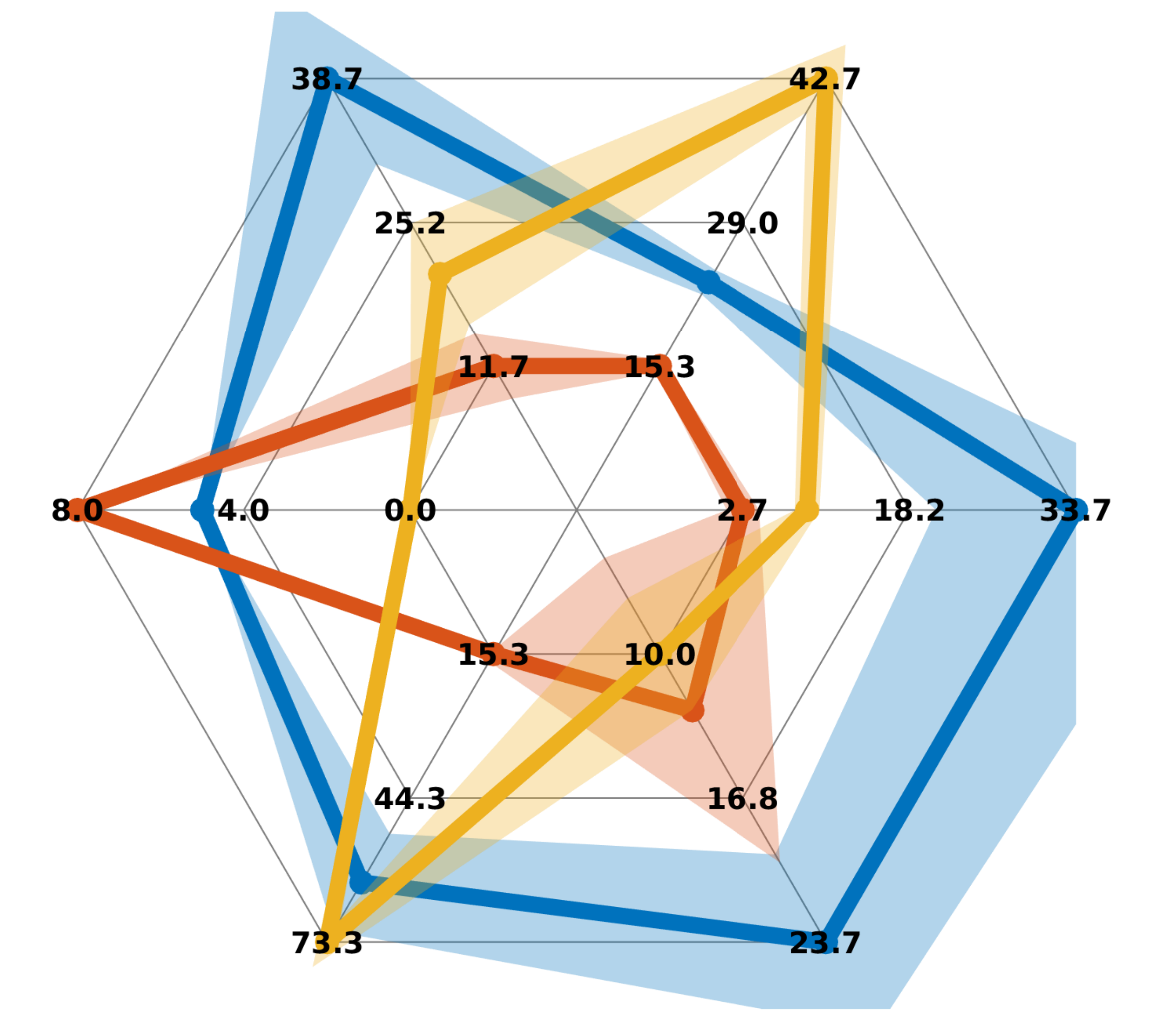}} \\ \hline
      
      \textbf{AO}  
      & \parbox[c]{1em}{\includegraphics[width=3.0cm]{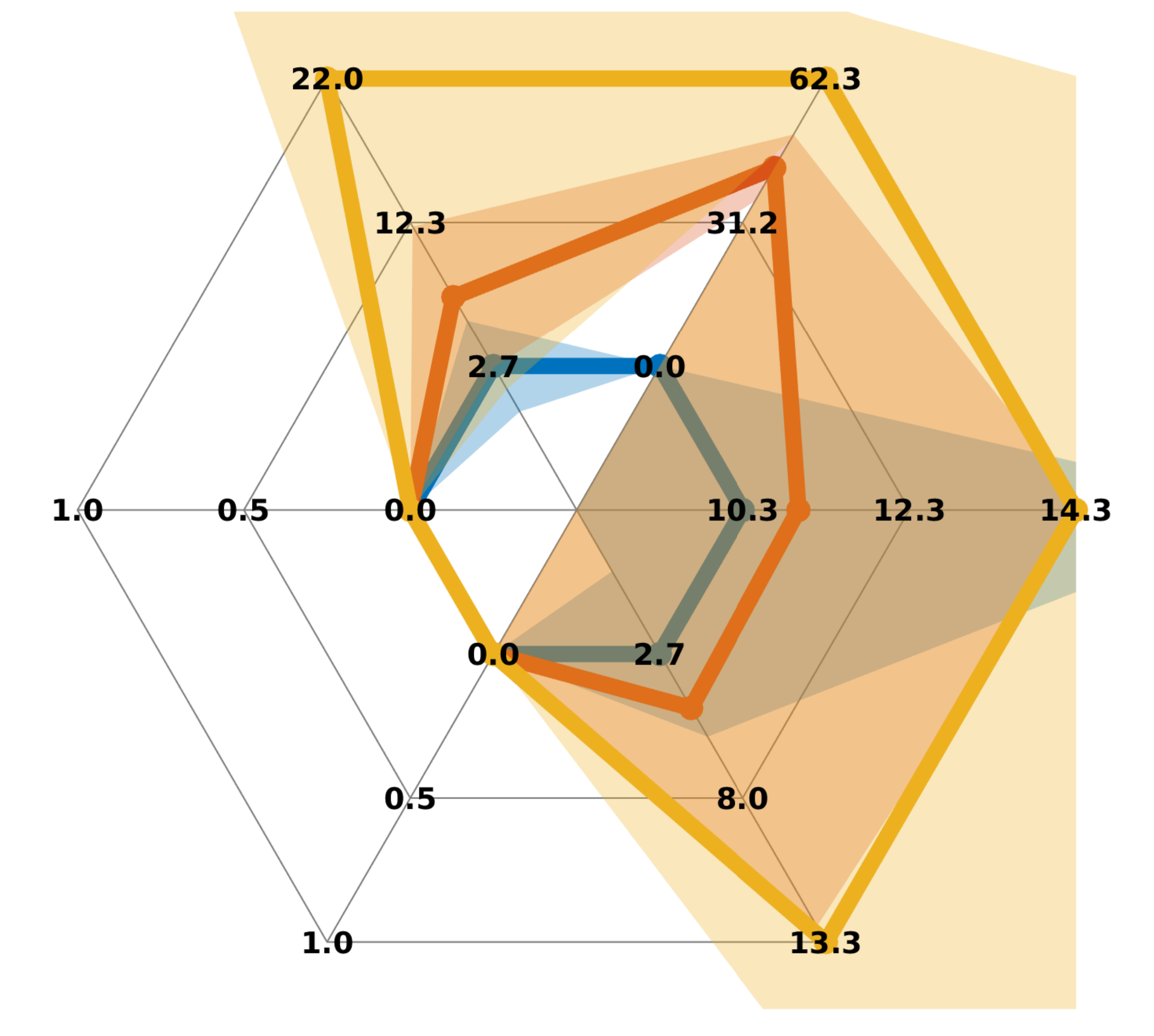}}     & 
      \parbox[c]{1em}{\includegraphics[width=3.0cm]{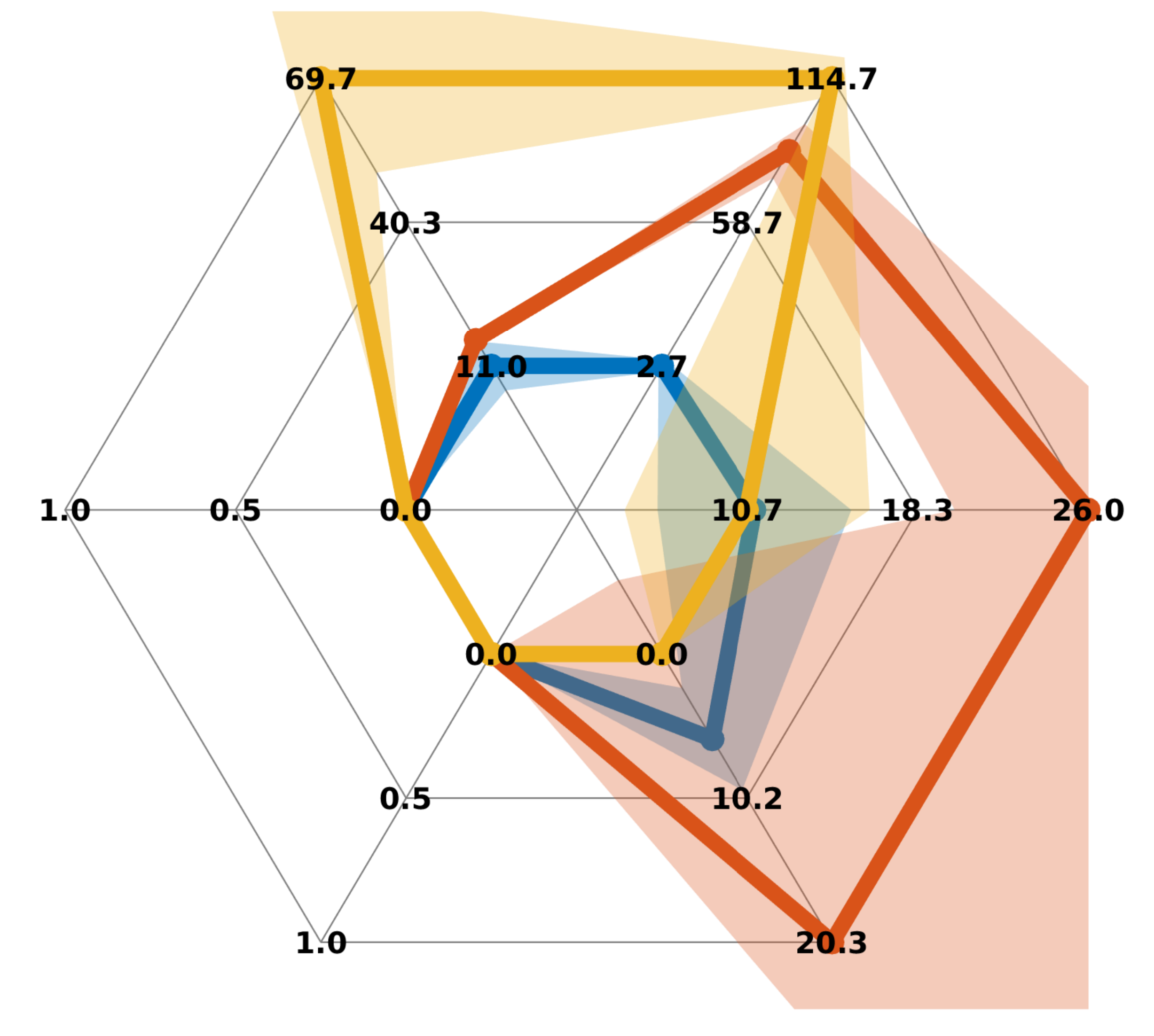}}  
      & \parbox[c]{1em}{\includegraphics[width=3.0cm]{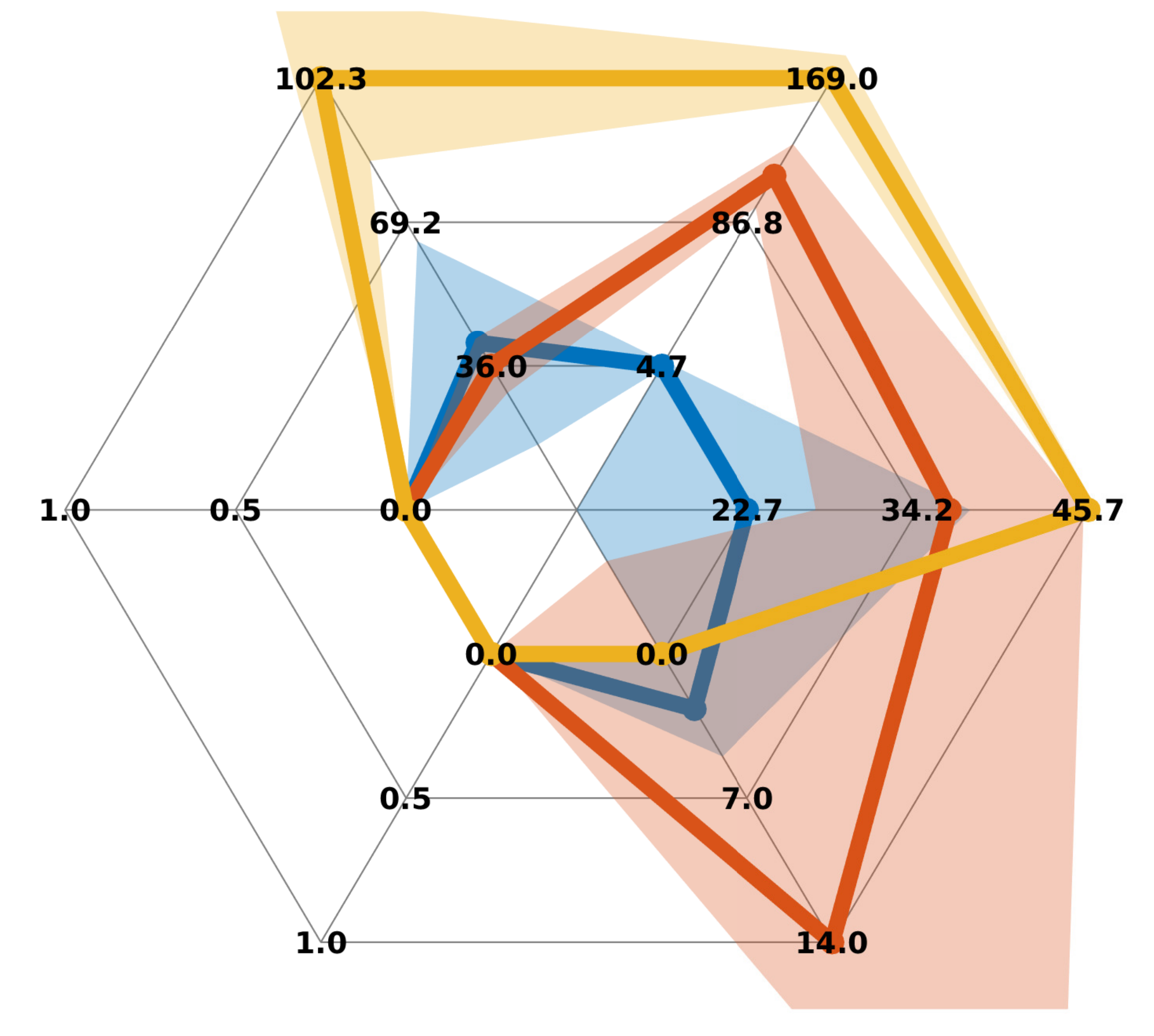}}  & 
      \parbox[c]{1em}{\includegraphics[width=3.0cm]{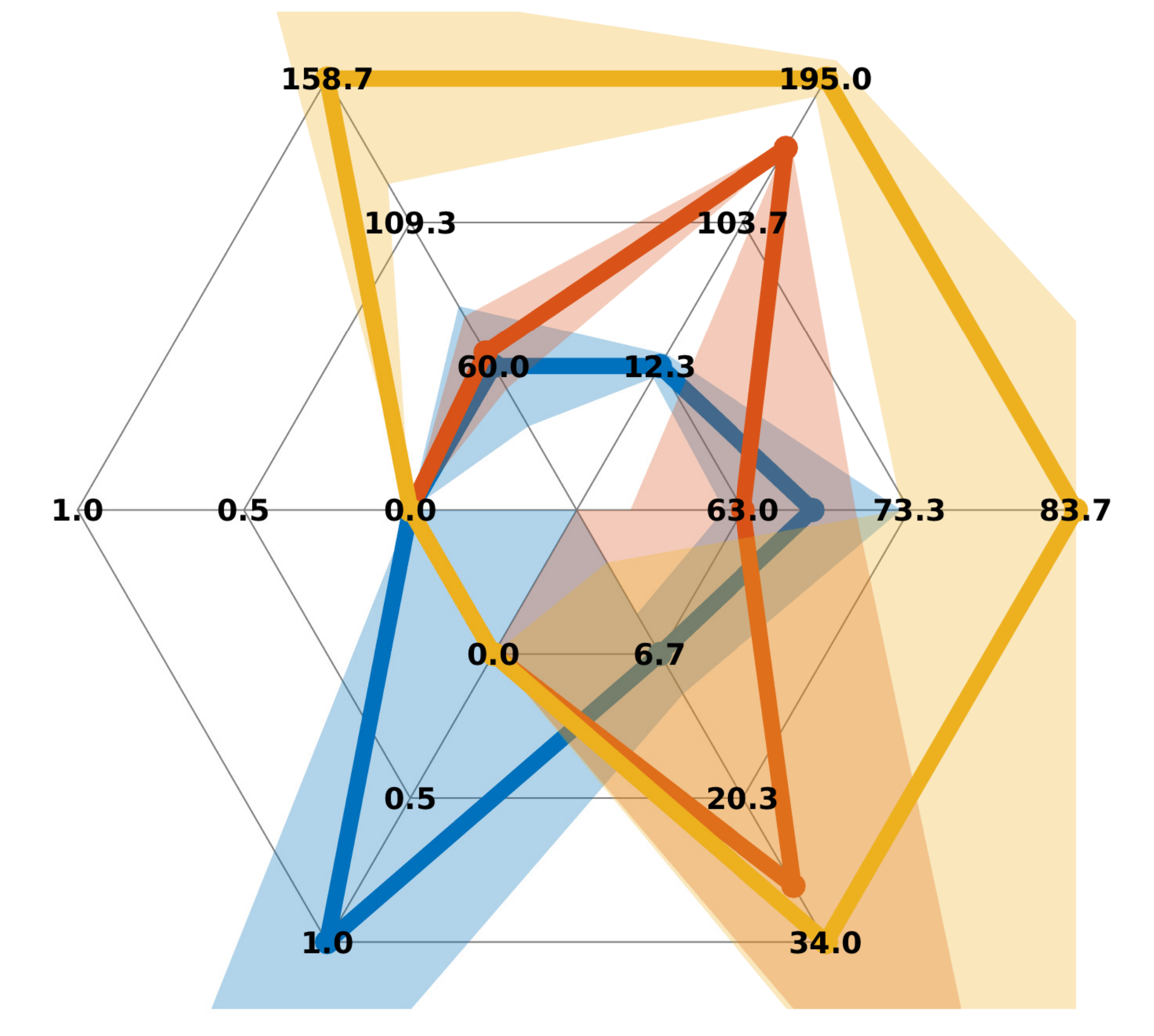}} 
      & \parbox[c]{1em}{\includegraphics[width=3.0cm]{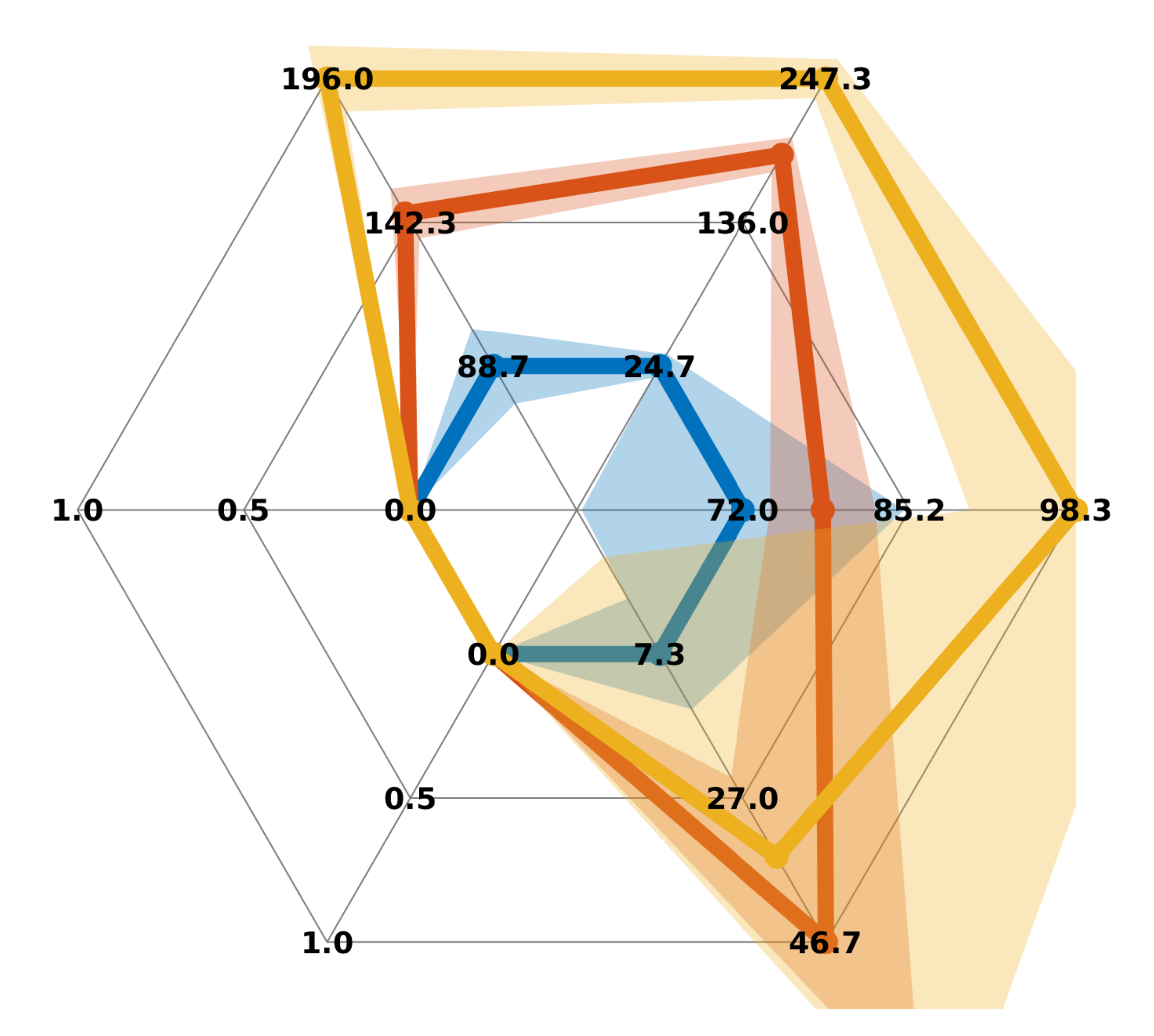}} \\ \hline
  \end{tabular}
\end{table*}

\subsection{Generalization to Test Scenarios}
Based on the above experimental setup, bidirectional experiments are conducted to test scenario generalization ability of the training paradigm-training domain combinations on test domains.\\

\noindent\textbf{Test on Exocentric Standard Scenarios} \\
In this test domain, models are evaluated on the six types of standard scenarios, varying in agent density from 10 to 50 and initial/destination positions. Fig.~\ref{fig:ranking_comparison_float} left shows the averaged rankings for the three metrics.

For DTW, \texttt{BCA-G}, \texttt{BCA-R}, \texttt{RLA-G} ranks first, second and third respectively. This indicates that BCA paradigm is better at inferring a route than RLA when the testing scenarios are widely divergent from the training scenarios. For AA, \texttt{BCA-G}, \texttt{BCA-R}, \texttt{RLA-G} ranks first, second and third respectively. For AO, surprisingly, \texttt{RLA-G} is the best while \texttt{BCA-G}, \texttt{BCA-R} ranks second and third respectively. Therefore. one can see that, under the same training paradigm (\texttt{BCA}), training on egocentric representative scenarios (\texttt{G}) incurs less AA and AO collisions than training on egocentric random scenarios (\texttt{R}), when applied to exocentric standard scenarios (\texttt{X}). This evidence that egocentric representative scenarios (\texttt{G}) provide a suite of challenging local agent-agent interactions and sufficient samples on avoiding collisions in a myriad of obstacle configurations. It also implies that when applying a model to a few challenging unseen environments (e.g., \texttt{X}), it might be better to train a model with a sufficient number of environment configurations (training on egocentric representative scenarios (\texttt{G}) enables the model to learn from 4000 different environments), than applying a model without environment knowledge (\texttt{BCA-R} learns from snapshots of surrounding neighboring agents, not from any specific environment).

To understand why \texttt{RLA-G} incurs less AO collisions than \texttt{BCA-G}, we further list Tab.~\ref{tab:spider_metrics_on_F} to show detailed comparisons along agent densities. We notice that the DTW metric of \texttt{RLA-G} is much higher than those of \texttt{BCA-G}. From simulation videos and trajectories illustrated in Fig.~\ref{fig:comparative_trajs_on_F}, we observe that only a few \texttt{RLA-G} agents can go through the doorway with slow speed, while other \texttt{RLA-G} agents have to wander near the doorway until the maximum number of simulation steps. The cautious behaviors of \texttt{RLA-G} agents bring benefits in terms of lower AO. This explains the outlier \texttt{RLA-G} in AO metric.\\


\noindent\textbf{Test on Egocentric Representative Scenarios}\\
In this evaluation, models are tested over 100 scenarios from the egocentric representative scenarios (\texttt{G}) domain. Fig.~\ref{fig:ranking_comparison_float} right shows averaged ranking results over three metrics. For DTW, \texttt{BCA-X}, \texttt{RLA-X}, \texttt{BCA-R} ranks first, second and third respectively. For AA, \texttt{BCA-R}, \texttt{BCA-X}, \texttt{RLA-X} ranks first, second and third respectively. For AO, again, \texttt{BCA-R}, \texttt{BCA-X}, \texttt{RLA-X} ranks first, second and third.

On the one hand, given the same training domain (exocentric standard scenarios (\texttt{X})), training with \texttt{BCA} paradigm is better than training with \texttt{RLA} paradigm for all metrics of DTW, AA, and AO. On the other hand, the training domain egocentric random scenarios (\texttt{R}) is better than exocentric standard scenarios (\texttt{X}), in term of reducing AA and AO collisions when generalized to many new scenarios. This implies an even more interesting insight: when a model needs to be applied to many new environments (testset of egocentric representative scenarios (\texttt{G}) comprises of 100 new environments), having no knowledge about any environment (\texttt{BCA-R}) is more advantageous than having a little knowledge about a few environments (\texttt{BCA-X} and \texttt{RLA-X} are trained only on six different environments).\\

\noindent\textbf{Overall Summary on Bidirectional Experiments}\\
According to the above bidirectional results and analysis, it is clear that \texttt{BCA} training paradigm is overall better than \texttt{RLA} training paradigm, and the data domain egocentric representative scenarios (\texttt{G}) and egocentric random scenarios (\texttt{R}) are better than exocentric standard scenarios (\texttt{X}) in reducing AA and AO collisions. Considering the coverage of these paradigms and domains, we conclude that (i) a simpler training paradigm is better than a more sophisticated training paradigm, (ii) training samples with diverse agent-agent and agent-obstacle interactions are beneficial for reducing collisions when the trained models are applied to new scenarios.

\subsection{Discussion} 
\label{sec8}

Results (Fig.~\ref{fig:ranking_comparison_float}) suggest that while \texttt{RLA}-based training methods have a potentially powerful paradigm of aggregate behavior imitation through a combination of IRL and RL, it may not possess the desired cross-domain generalization observed in a simpler \texttt{BCA} paradigm, provided that all models have the same architecture and the same number of parameters. One reason for this may stem from the underlying modeling assumptions.

As evident from the expression of occupancy measure, \texttt{RLA} relies on matching the occupancy measures between the estimated policy $\pi^{*}$ and the expert policy $\pi_{E}$. \cite{puterman2014markov} shows that a valid set of occupancy measures $\mathcal{D} \triangleq \{ \rho_{\pi} | \pi \in \Pi \}$ satisfies a set of affine constraints:
$\sum_{a} \rho(s,a) = p_{0}(s) + \gamma \sum_{(s', a)} P(s | s', a) \rho(s', a), \forall s \in \mathcal{S}$, where $p_{0}(s)$ denotes the distribution of initial states. Moreover, there is a one-to-one correspondence between $\mathcal{D}$ and $\Pi$: 
$\pi_{\rho}(a|s) \triangleq \frac{\rho(s,a)}{\sum_{a'}\rho(s,a')}$,
with $\pi_{\rho}$ the unique policy whose occupancy measure is $\rho$, Thm.2 of~\cite{syed2008apprenticeship}. Taking this into account, we obtain:
\begin{align}
\label{eq:policy_encode_dynamics}
\pi_{\rho}(a|s) = \frac{\rho(s,a)}{p_{0}(s) + \gamma \sum_{(s', a)} P(s | s', a) \rho(s', a)}, \forall s \in \mathcal{S}. \nonumber
\end{align}

Thus, when modeling the movement of agents in an environment, the dynamics $P(s|s', a)$ encodes complex scenario information, including positions of other moving agents and the obstacles in the environment, occlusions, etc. These dynamics are, as noted, implicitly encoded in the policy. Therefore, an \texttt{RLA} model trained on a particular training domain implicitly learns its environments. Transferring this model directly to a new, test scenario with significantly different dynamics is bound to result in a weaker match, thus reduced generalization capacity. On the other hand, less biased \texttt{BCA} models will have the ability to surmount those differences more easily, and generalize better.

\section{Generalization to Real Domain}
\label{generalization_to_real_domain}

In this section, we apply the above five combinations of training paradigms and training domains to a real test domain to visualize their scenario generalization abilities and verify the conclusion in a real world domain.

\subsection{Real Domain Description}
The real domain we considered is Stanford crowd trajectory dataset, introduced in~\cite{alahi2014socially}. It consists of a large set of real pedestrian trajectories collected at a train station of size $25 \text{m} \times 100 \text{m}$ for $12 \times 2$ hours by a set of distributed cameras. Identity numbers, position histories with timestamps of the pedestrians are extracted from the image sequences with detection and tracking algorithms. The dataset is challenging since (1) The agent density is quite high. In a time duration of 4 minutes, there are about 500 pedestrians moving in the train station. (2) Pedestrians are highly asynchronous. They enter into and exit from the train station at different timestamps, without a unified time controller. (3) the data is noisy, due to the detection, tracking and localization error, and the difficulty to measure the accurate positions of the obstacles (infeasible areas).

\subsection{Dataset Preprocessing}
First, the positions of the obstacles in the environment layout are determined by drawing the provided pedestrians' trajectories on the layout, and manually finding out the obstacle positions based on the occupancy areas of the drawn trajectories. Second, the long-lasting trajectories are aligned with timestamps and further split with a temporal sliding window of 4-minute length and 2-minute stride. Within each time window, all pedestrians are retrieved, including those emerge after the starting time and/or exit before the ending time of the window, and those whose destinations have to be retrieved in the next time window. Third, to reduce the noise in the data, pedestrians whose initial position or destination position are within the obstacles are removed. Last but not the least, a Gaussian convolution operation is applied to the binary representation of the environment layout (obstacle pixels are represented as 1, other feasible pixels are 0), to yield an obstacle-probability map. Based on the map, the cost from a node to its child node in A-star is modified according to the obstacle probability, so as to prevent the planned A-star nodes from being too close to the obstacles, to reduce the risk of agent-obstacle collisions.

\subsection{Visualization of Model Trajectories}
\begin{figure}[t]
 \begin{subfigure}{0.48\textwidth}
 \centering
 \includegraphics[width = \textwidth]{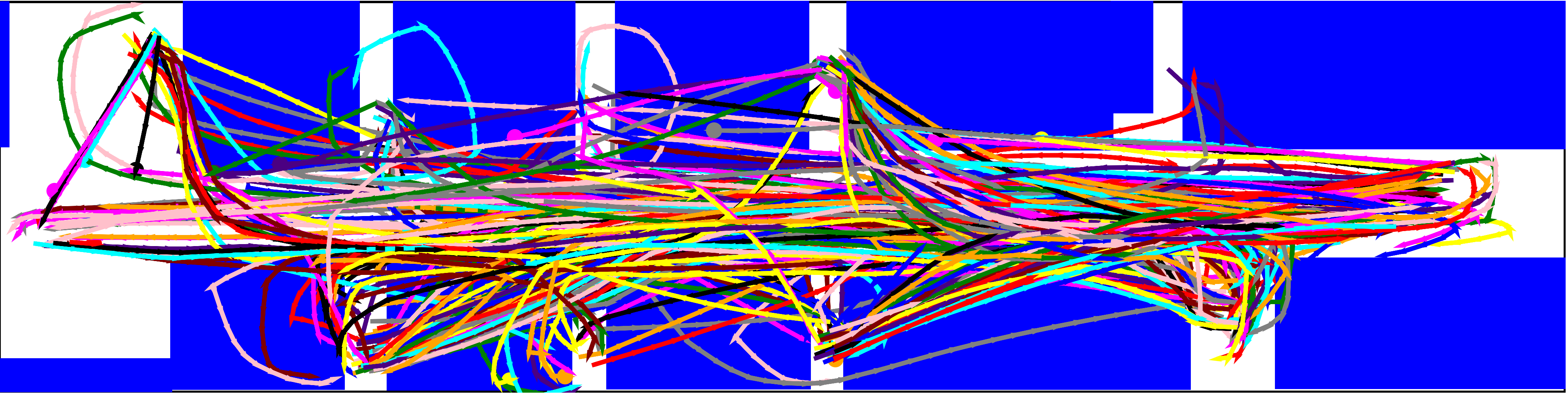}\\
 (1) \texttt{RLA-X} in a time window \vspace{3mm}\\
 \end{subfigure} 
 \begin{subfigure}{0.48\textwidth} 
 \centering
 \includegraphics[width = \textwidth]{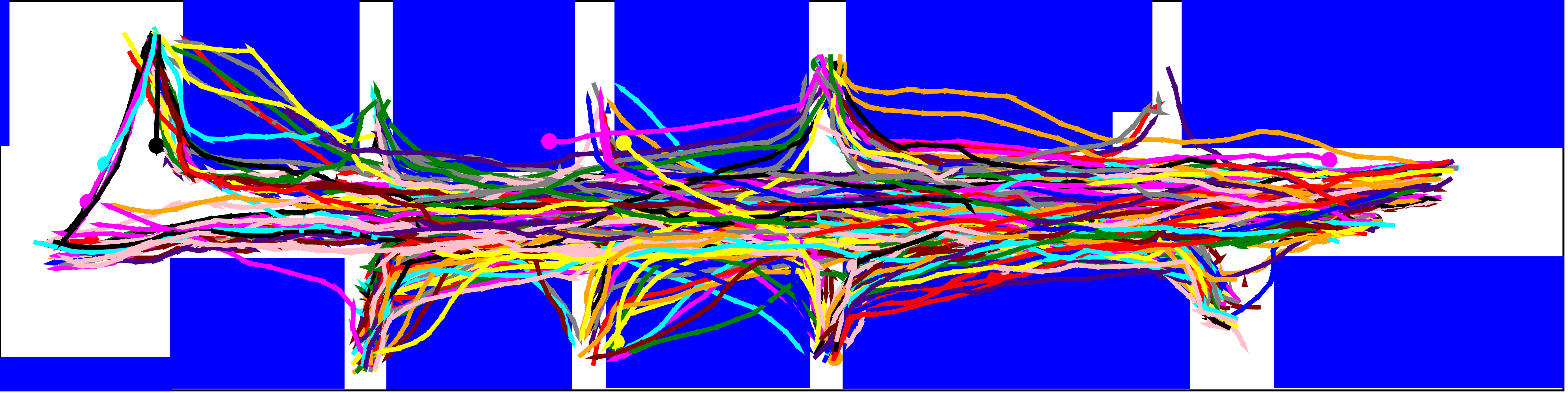}\\
 (2) \texttt{RLA-G} in a time window \vspace{3mm}\\
 \end{subfigure} 
 \begin{subfigure}{0.48\textwidth}
 \centering
 \includegraphics[width = \textwidth]{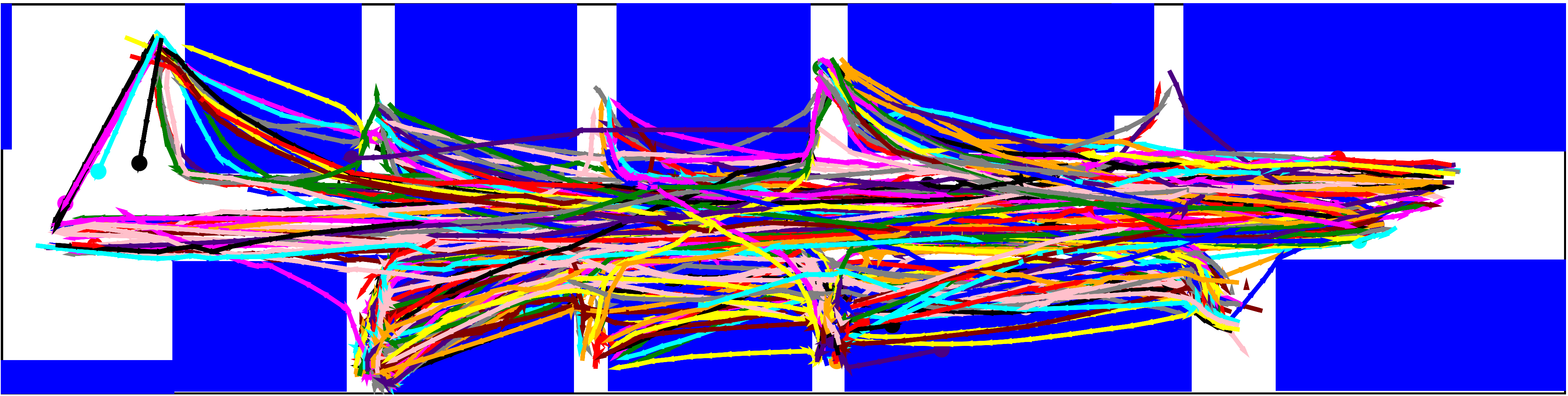}\\
 (3) \texttt{BCA-X} in a time window \vspace{3mm}\\
 \end{subfigure} 
 \begin{subfigure}{0.48\textwidth}
 \centering
 \includegraphics[width = \textwidth]{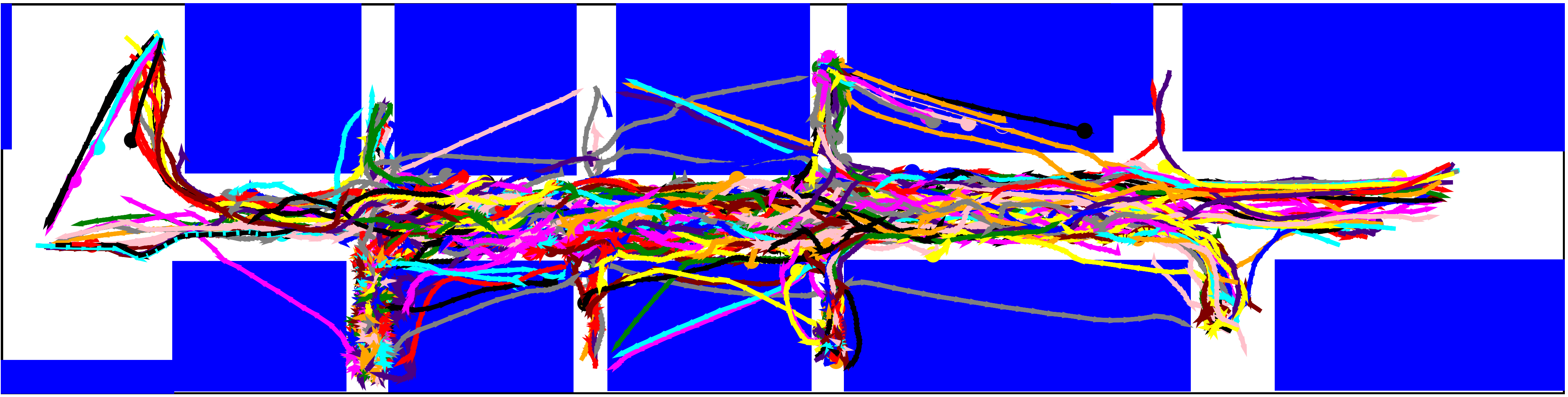}\\
 (4) \texttt{BCA-G} in a time window \vspace{3mm}\\
 \end{subfigure}
 \begin{subfigure}{0.48\textwidth}
 \centering
 \includegraphics[width = \textwidth]{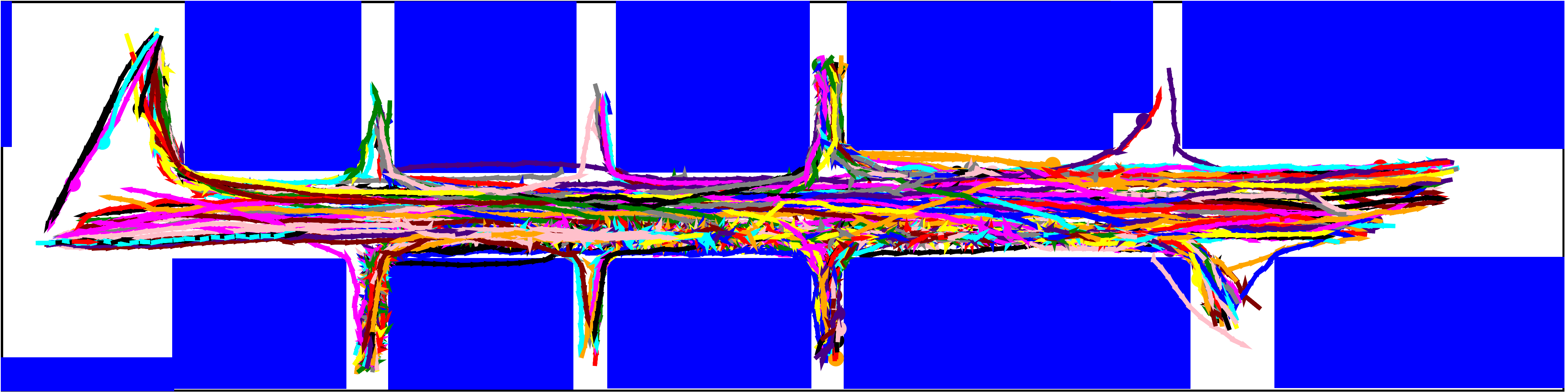}\\
 (5) \texttt{BCA-R} in a time window
 \end{subfigure}
\caption{Visualization of different combinations of training methods and training domains generalized to real dataset. The obstacles (infeasible areas) are in blue color.}
\label{fig:real-domain}
\end{figure}

Fig.~\ref{fig:real-domain} illustrates trajectories of the above five combinations of training methods and training domains on the Stanford real dataset. The obstacles (infeasible areas) are in blue color, and the trajectories are also colored. According to our experiment setting, we know that for agents even slightly entering into an obstacle, they will not perceive the obstacle wherein. However, in this specific test domain, some agents slightly entering into the obstacle may see their far-away planned nodes (e.g., the final destination node), and thus would be guided to directly approach to their final destinations, leading to visually obvious agent-obstacle collisions. We can see that even under such challenging scenarios, with high agent density and easy-to-cause obstacle crossing, \texttt{BCA-R} and \texttt{BCA-G} are still visually generalized better than other combinations. Thus the visualization strengthens our conclusion.

\subsection{Quantitative Results}
\begin{figure}[htb]
 \centering
 \includegraphics[width = 0.42\textwidth]{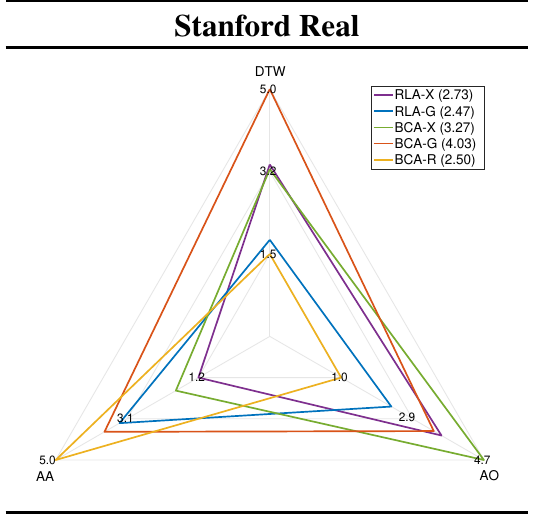}
 \caption{\small Rankings of all models over real test domain. Each color represents a model. Each axis indicates averaged rankings over test scenarios for a metric. All metrics are the smaller the better. Numbers in legend denote averaged ranking over three metrics.}
\label{fig:ranking_on_real}
\end{figure}

Fig.~\ref{fig:ranking_on_real} presents the averaged rankings of all models when generalized to the real domain on the three metrics. We can see that for DTW, \texttt{BCA-R} and \texttt{RLA-G} ranks first and second respectively. For AA, \texttt{RLA-X} and \texttt{BCA-X} ranks first and second respectively. For AO, \texttt{BCA-R} and \texttt{RLA-G} ranks first and second respectively. Overall, \texttt{RLA-G} and \texttt{BCA-R} models are better than others.

From the rankings we have three observations. (1) Training domain egocentric random (\texttt{R}) and training domain egocentric representative (\texttt{G}) are beneficial for reducing AO collisions, which accords with the simulated bidirectional experiments. (2) Training domain exocentric standard (\texttt{X}) is better at reducing AA collisions. This suggests that even though the exocentric standard (\texttt{X}) domain is not suggested by the simulated bidirectional experiments, it contains a few challenging obstacle configurations and can still benefit a model when applied to real challenging scenarios with high agent density. (3) For the training paradigm, both \texttt{RLA} and \texttt{BCA} are involved in the first and second ranked models in each of the three metrics and in the overall ranking. The lack of a dominant training paradigm implies the need to trade off when choosing a training paradigm for generalizing to real challenging domains.

\section{Conclusion}

In this study, our main goal is to analyze the effect of different training paradigms and training domain characteristics on scenario generalization capacities of data-driven imitation models in crowd modeling settings. Our empirical results and analysis indicate that for training method, the simpler behavior cloning method is overall better than the more complex reinforcement learning method. According to our experiment results, it is also noticeable that the training domains have substantial impact on the generalization ability of models to new scenarios. In particular, training samples with diverse agent-agent and agent-obstacle interactions are beneficial for reducing collisions when models are applied to new scenarios. 

Future work includes: (1) a comparison to scenario generalization capacities of RL agents whose reward functions are pre-defined, for example, as a combination of the three metrics (DTW, AA, AO); (2) the improvement of scenario generalization capacity. For instance, train a model in a training domain and then adopt it in a testing domain using limited testing samples ~\cite{zhao2018adversarial,le2018theoretical}, where a domain is the dynamics belonging to a specific type of scenarios.\\

\section*{Acknowledgments}  
Kapadia has been funded in part by NSF IIS-1703883, and NSF S\&AS-1723869. Yoon has been funded in part by TCNJ SOSA 2017-2019 grant.


\bibliographystyle{aaai.bst}
\bibliography{publist}  
\end{document}